\documentstyle[12pt,epsf]{article}
\def\bm#1{\mbox{\boldmath $#1$}}
\def\it#1{\mbox{$#1$}}

\def\beq{\begin{equation}}
\def\eeq{\end{equation}}

\def\t2{\mbox{  }}
\def\rst1{\mbox{ }}

\def\frc#1,#2{{{#1}\over{#2}}}
\begin{document}
\setcounter{equation}{0}
\renewcommand{\theequation}{\arabic{equation}}
\renewcommand{\thesection}{\Roman{section}.}
\renewcommand{\thesubsection}{\Alph{subsection}.}
\renewcommand{\thesubsubsection}{\Alph{subsection}.\arabic{subsubsection}.}

\title{Lekner summations and Ewald summations for quasi-two dimensional
systems.}    
\author{\normalsize Martial Mazars\footnote{Electronic mail: Martial.Mazars@th.u-psud.fr}\\
\small Laboratoire de Physique Th\'eorique (UMR 8627),\\
\small Universit\'e de Paris XI, B\^atiment 210, 91405 Orsay Cedex,
FRANCE}
\maketitle
\hfill\small\hspace{3.0in} L.P.T.-Orsay 	\\

\begin{center}{\bf Abstract}\end{center}
Using the specific model of a bilayer of classical charged particles
(bilayer Wigner crystal), we compare the predictions for energies and
pair distribution functions obtained by Monte Carlo simulations using
three different methods available to treat the long range Coulomb
interactions in systems periodic in two directions but bound in the
third one. The three methods compared are: the Ewald method for
quasi-two dimensional systems [D.E. Parry, Surf. Sci. $\bm{49}$, 433
(1975); \it{ibid.}, $\bm{54}$, 195 (1976)], the Hautman-Klein method
[J. Hautman and M.L. Klein, Mol. Phys. $\bm{75}$, 379 (1992)] and the
Lekner summations method [J. Lekner, Physica A$\bm{176}$, 485
(1991)]. All of the three method studied in this paper may be applied to
any quasi-two dimensional systems, including those having not the
specific symmetry of slab systems. For the particular system used in
this work, the Ewald method for quasi-two dimensional systems is exact
and may be implemented with efficiency; results obtained with the other
two methods are systematically compared to results found with the Ewald
method. General recommendations to implement with accuracy, but not
always with efficiency, the Lekner summations technique in Monte Carlo
algorithms are given.      

\newpage

\section{Introduction.}

In Chemical physics, Condensed matter physics and Molecular physics,
numerical simulations are basic theoretical tools to study generic model
systems and to predict qualitatively and/or quantitatively the
properties of some materials. Molecular simulations allow to study the
thermodynamical and the transport properties of some more or less
detailled models of real substances. In Molecular simulations, there are
mainly two procedures used: the Molecular Dynamics method and Monte
Carlo algorithms, both being generally considered as
complementary \cite{r1,r2}.\\   
The number of particles or the number of degree of freedom of a system
generally considered in these studies is ranging from few hundreds to
few millions. Compared to macroscopic materials, this number is small
and far from the thermodynamical limit. To avoid irrelevant surface or
finite-size effects, periodic boundary conditions are generally used.\\ 
When the interaction is short ranged, truncations of potentials or
minimum image conventions are used to compute with accuracy the
interaction energy or forces between molecules or particles.\\
On the other hand, when the interaction is long-ranged, as it is the case
for Coulombic and Dipolar interactions, truncation of the potential may
lead to severe bias \cite{r3}. For bulk-like systems with Coulombic or
Dipolar interactions, Ewald methods are employed to take into account
the long-ranged contribution of the potential to energy or forces when
periodic boundary conditions are applied to the system [4-6].\\      
In the context of molecular simulations, the interesting feature of
Ewald summations method for bulk-like systems is that the cost in
computing time for the calculation of the energy of a configuration may
scales as $N^{\frac{3}{2}}$, where $N$ is the number of particles. For
large systems, this computational time is still too expensive and some
numerical procedures based on the Fast Fourier Transform algorithm like
the particle-particle/particle-mesh (PPPM) method \cite{r7}, or procedures
based on multipole expansion like the fast multipole methods \cite{r8} have
been implemented. For bulk-like systems, all these methods have been
used and conveniently intercompared in the last two decades.\\  
Some very interesting systems may not be considered as bulk-like
systems, for instance when one or two dimensions of the system are small
compared to the other dimensions: the quasi-one dimensional or quasi-two
dimensional systems. Examples of such systems are fluid-fluid or
fluid-solid interfaces, monolayers, biological membranes, free standing
films, cylindrical pores, carbon nanotubes, etc. As for bulk-like
systems, to avoid some irrelevant finite size effects, other than the
finite extension of these systems, partially periodic boundary
conditions are employed. Partially periodic boundary conditions consist
of taking periodic boundary conditions for only one or two dimensions,
while the remaining dimensions are considered with their finite
extension. For these systems, straightforward applications of the Ewald
summations lead to a computational effort that scales as $N^{2}$ and not
as $N^{3/2}$, because the analytical formulation of the electrostatic
energy of this kind of systems is more complicated than for bulk-like
systems [9-15]. To compute coulombic or dipolar energies in these
quasi-two (or quasi-one) dimensional systems, several others methods have
been proposed [9,16-23]. One has to note that some of these methods
may be applied only to systems having a slab-like geometry, while the
methods compared in this paper may be applied to any quasi-two
dimensional systems.\\  
The purposes of the present work is to exhibit a comparison between three
different methods proposed to study quasi-two dimensional systems. The
method compared in this work are: (a) the Ewald method for quasi-two
dimensional systems (Ewald quasi-2D) [10,11,13-15], (b) the
Hautman-Klein method \cite{r16} and (c) the Lekner summations
technique \cite{r19}. All the three methods are applied to a very simple
system: a bilayer Wigner crystal, described in the section II of this
paper. As it is explained further in the paper, we have chosen to apply
these methods to this particular system because for this system the
Ewald quasi-2D method may be implemented with the same efficiency as the
Ewald method for bulk-like systems. To make consistent intercomparisons
between methods, one needs to have some reliable and solid reference
points, the Ewald quasi-2D method is the best candidate to produce such
reference points. All the results obtained by applying the Hautman-Klein
method and the Lekner summations technique to this system are
systematically compared to the results obtained by applying the Ewald
quasi-2D method.\\  
This work follows a previous study on the convergence of Lekner
summations done by the author in ref.[24] and another study done in
ref.[25] where we have compared the methods of references [16,18,20] on
a system having a slab geometry. One of the aims of the present work is
also to present explicitly a procedure to implement with accuracy (but
not always with efficiency) the Lekner summations technique in
Monte-Carlo simulations using the Metropolis algorithms.\\
Comparisons of Ewald summation techniques for confined and
quasi-bidimensional systems have been done previously, such as in the
interesting work by A.H. Widmann and D.B. Adolf \cite{r26}. In these
studies, authors generally compare numerical accuracy and efficiency of
methods on only few configurations of a simple system; for instance in
the study of ref.[26], Widmann and Adolf compared Hautmann-Klein,
Nijboer-de Wette and Parry methods on five configurations of a system
with 100 particles. In the present work, the Ewald quasi-2D,
Hautmann-Klein and Lekner methods are implemented in Monte Carlo
simulations; the influence of the accuracy on thermodynamical averages
and on correlation functions is outlined. The purpose of the present
paper is to concretely illustrate, on a very simple system, how some
methods, and especially the Lekner method, may introduce biais on
physical quantities in a Monte Carlo simulation, since relations between
thermodynamical averages and accuracy of methods are needed to validate
outputs of numerical simulations.\\         
In section II, we describe the model on which comparisons between the
different methods are made. In section III, we present the three methods
that are studied and compared in this paper. Section IV is devoted to
exhibit the results obtained with Monte Carlo simulations using the
Metropolis algorithm. A general discussion and some recommendations to
implement correctly and with accuracy the Lekner summations are given in
section V.\\  
For completeness, an appendix that gives the computation of the
interaction between two charged surfaces and ions has been added at the
end of the paper.\\  

\section{Description of the Model and simulations parameters.} 

A bilayer of charged particles is the 'reference' system chosen in this
work. The three methods used to compute the electrostatic energy of this
system are presented in the next section.\\
This model is frequently used to give crude representations of very
different systems. For instance such bilayer Wigner crystals are used to
give representation of strongly coupled electronic bilayers of charged
particles in two dimensional semiconductor heterostructures or in dusty
plasmas. Some recent experiments on Laser-cooled $^{9}$Be$^{+}$ ions
have provided direct observations of structural phase transition in
these systems \cite{r27}.\\      
This very simple model is also used to give very crude representation of
neutral lipid bilayers \cite{r28} or as basic theoretical model to
facilitate the conceptual understanding of counterion-mediated
attractions between similarly charged planes \cite{r29}. The monolayer
version of this model is also useful to describe classical electrons
trapped on the surface of liquid helium \cite{r30}.\\ 
Our reference system is a bilayer made of $N=2N_{0}$ point ions
interacting by a Coulomb potential $1/r$. The ions are evenly
distributed in two layers $\L_{1}$ and $\L_{2}$ separated by a distance
$h$. A snapshot of an instantaneous configuration of a system of 512
point ions is represented on Figure 1. Each layer is a square of side
$L$ and partial periodic boundary conditions, with a spatial periodicity
$L$, are applied in both directions parallel to the layer (directions
$x$ and $y$) while no periodic boundary conditions are taken in the
third direction (direction $z$). The purpose of this work is not to
study the phase diagram of this system (which may be very rich, see for
instance ref.[31]), thus in all computations that we have done, the
shape and area of the layers are constants; nevertheless most of the
computations done in this paper can be compared to a more extensive
thermodynamical study of this system done in ref.[32]. The charge of
points ions is $q$ and to guaranteed charge neutrality, both layers
$\L_{1}$ and $\L_{2}$ carry a uniform surface charge density $\sigma$
given by      
\beq
\displaystyle Nq+2\sigma L^{2}=0 
\eeq
In all computations done, we have taken $q=14$. The characteristic
length $a$, the ion-disk radius, is defined by $\pi\rho_{0}a^{2}=1$
where the ion density in each layer is $\rho_{0}=N_{0}/L^{2}$. This
ion-disk picture is a simplified version of the hexagonal Wigner-Seitz
cell.\\    
We have made computations using the three differents methods described
in the next section in six differents situations resumed in TABLE I. For
Runs a, c, d, e, e' and f, the coupling constant is
$\Gamma=q^{2}/kTa=196$ ($a=1$), while for Runs b, it is $\Gamma=98$
($a=2$). For all computations, a Monte Carlo cycle (MC-cycle) is made of 
a random trial move of each particle in both layers, the trial moves are
accepted or rejected according to the Metropolis algorithm applied to
the NVT ensemble \cite{r33,r1,r2}; in a MC-cycle there are $N$ trial moves. The
amplitude of the trial displacement has been chosen such as the
acceptance ratios ranged between $30\%$ and $60\%$. No exchange of ions
between layers $\L_{1}$ and $\L_{2}$ is allowed. From an initial
configuration, $t_{eq}$ MC-cycles are used for equilibration and after
$t_{av}$ MC-cycles are used to accumulate the thermodynamical
averages. The differences between Runs e and e' are in the values of
$t_{eq}$ and $t_{av}$; Runs e' were applied only to computations using
the Ewald quasi-2D and Hautman-Klein methods.\\    
In Runs a, c, d, e, e' and f, the thermodynamical stable state is a
square or hexagonal solid; for these computations, the average
electrostatic energies per particle are closed to the corresponding
Mandelung energies. The Mandelung energy, $\beta u_{0}$, for a square
bidimensional lattice may be found in a computation done by
Totsuji \cite{r5}. Taking the lattice length as $L/\sqrt{N_{0}}$, the
Mandelung energies computed in the work of Totsuji are given in TABLE I
for all Runs done in the present work. Since the shape and the area of
the bilayer are kept constant, this would impose preferentially a square
lattice for the solid-like phase, even if the thermodynamical stable
phase has another symmetry. To study rigourously the solid phase diagram
of this bilayer system, one should allow at least the shape of the box
to fluctuate. This study was done in ref.[32]. In the present work, the
shape, a square of side $L$, and the area $L^{2}$ of layers are constant
and the average energy per particle found in Monte Carlo computations,
except for Runs b, is closed to $u_{0}$. We also give in TABLE I, the
preferential structures found in the computations done in ref.[32] for
the bilayer Wigner crystal.\\   
The order parameters $\Psi_{m}$ used in ref.[32] to differentiate the
square and hexagonal structures are not computed in the present
study. In the spirit of the current work, these parameters are
irrelevant and since the fixed square shape of layers would induced
preferentially a square lattice for the solid-like phases, the values
for $\Psi_{4}$ and $\Psi_{6}$ that would have been found might be
misleading to further studies of the solid phase diagram of this bilayer
system.\\   
The energy of the bilayer system can be written as
\beq
U=E_{intra}+E_{inter}
\eeq
where $E_{intra}$ is the intralayer contribution and $E_{inter}$ is the
interlayer contribution, including the contribution of the
surface-surface energy (see appendix). In the next section, we give
the analytical expressions for $E_{intra}$ and $E_{inter}$ for each of
the methods studied in this work.\\
The intralayer $g_{11}$ and interlayer $g_{12}$ pair distribution
functions, corresponding to particles in the same layer and to particles
belonging to differents layers but positions projected onto the same
plane respectively, are evaluated. These distribution functions are
computed as  
\beq
\left\{\begin{array}{ll}
\displaystyle g_{11}(s) &\displaystyle = \frac{L^{2}}{2N_{0}^{2}}<\sum_{i\in\L_{1}}\sum_{j\in\L_{1},j\neq i}\delta(\bm{s}-\bm{s}_{ij})+\sum_{i\in\L_{2}}\sum_{j\in\L_{2},j\neq i}\delta(\bm{s}-\bm{s}_{ij})>\\
&\\
\displaystyle g_{12}(s) &\displaystyle = \frac{L^{2}}{N_{0}^{2}}<\sum_{i\in\L_{1}}\sum_{j\in\L_{2}}\delta(\bm{s}-\bm{s}_{ij})>
\end{array}
\right.
\eeq
where particles positions are noted by
$\bm{r}_{i}=\bm{s}_{i}+z_{i}\bm{e}_{z}$ and $<\dots>$ denotes
statistical averages.\\  
In the next section, we describe the different methods used to compute
the electrostatic energy, while in section IV we present the results
obtained by applying these methods to the bilayer system. 

\section{Numerical methods to compute the electrostatic energy.} 

An efficient technique for computing the long ranged part of the
intermolecular interactions for systems with periodic boundary
conditions is provided by Ewald sums \cite{r1,r2,r4}. In this procedure, the
interaction energy between molecules is separated into short ranged and
long ranged contributions by using a damping function $f(r;\alpha)$,
where $\alpha$ is a convergence parameter conveniently chosen. For the
coulombic interaction, this separation is done by setting   
\beq
\frac{1}{r}= \frac{1-f(r;\alpha)}{r}+\frac{f(r;\alpha)}{r}
\eeq
The short range contributions to the energy are given by the first term
of the right handed side (r.h.s) of Eq.(1) and the long ranged
contributions are given by the second term. To handle the long ranged
contributions, Fourier transform of the second term is taken; the
interaction energy can be set in the form
\beq
E=E_{\bm{r}}^{(s)}+E_{\bm{k}}^{(l)}
\eeq
where $E_{\bm{r}}^{(s)}$ is the short ranged contribution and
$E_{\bm{k}}^{(l)}$ the long ranged contribution, which for technical
convenience, is expressed as  
\beq
E_{\bm{k}}^{(l)}=E_{\bm{k}=\bm{0}}^{(l)}+E_{\bm{k}\neq\bm{0}}^{(l)}
\eeq
The damping function $f(r;\alpha)$ may be chosen such as both
contributions of Eq.(5) are rapidly convergent. Such choices allow to
derive efficient algorithm to compute the interaction energy.\\ 
The choice 
\begin{center}
$\displaystyle f(r;\alpha)=\mbox{erf}(\alpha r)=\frac{2}{\sqrt{\pi}}\int_{0}^{\alpha r}dt\rst1\exp(-t^{2})$ 
\end{center}
is frequently used. Other derivations using an integral representation
of the Gamma function give the same damping function \cite{r13,r15}. We found
for the Coulomb interaction energy of $N$ particles in a simulation box
of side $L$ with periodic boundary conditions in the three directions, 
\beq
\left\{\begin{array}{ll}
&\displaystyle E_{\bm{r}}^{(s)}=\frac{1}{2}\sum_{i=1}^{N}\sum_{j=1}^{N}\sum_{\bm{n}}'q_{i}q_{j}\frac{\mbox{erfc}(\alpha|\bm{r_{ij}}+\bm{L}_{\bm{n}}|)}{|\bm{r_{ij}}+\bm{L}_{\bm{n}}|}\\
&\\
&\displaystyle E_{\bm{k}=\bm{0}}^{(l)}=\frac{2\pi}{3V}\mbox{\large (}\sum_{i=1}^{N}q_{i}\bm{r}_{i}\mbox{\large )}^{2}-\frac{\alpha}{\sqrt{\pi}}\sum_{i=1}^{N}q_{i}^2 \\\\
&\\
&\displaystyle E_{\bm{k}\neq\bm{0}}^{(l)}=\frac{2\pi}{V}\sum_{\bm{k}\neq\bm{0}}\frac{\exp(-k^{2}/4\alpha^{2})}{k^2}\mbox{\large $|$}\sum_{j=1}^{N}q_{j}\exp(i\bm{k}.\bm{r}_{j})\mbox{\large $|$}^{2}
\end{array}
\right.
\eeq
with
\begin{center}
$\displaystyle \bm{L}_{\bm{n}}=n_{x}L_{x}\hat{\bm{e}}_{x}+n_{y}L_{y}\hat{\bm{e}}_{y}+n_{z}L_{z}\hat{\bm{e}}_{z}$
\end{center}
where $n_{x}$, $n_{y}$ and $n_{z}$ are relatives integers.\\
The primed sum in $E_{\bm{r}}^{(s)}$ indicates that for
$\bm{L}_{\bm{n}}=\bm{0}$, $i=j$ must be omitted. In Eq.(7),
$\bm{r}_{i}=(x_{i},y_{i},z_{i})$ are positions of particles,
$\bm{r}_{ij}=\bm{r}_{i}-\bm{r}_{j}$ and $\bm{k}$ the lattice vectors in
the tridimensional Fourier space.\\    
In Eqs.(7), the long ranged contributions $E_{\bm{k}\neq\bm{0}}^{(l)}$
are expressed with a summation over the particles, $N$ contributions to
the sum, and not as a summation over the pair of particles, $N(N-1)/2$
contributions; this factorization is one of the technical features that
allows to implement efficiently the Ewald-3D method. By reference to the
crystalographic origin of the Ewald summations,
$\tilde{\rho}(\bm{k})=\sum_{j}q_{j}\exp(i\bm{k}.\bm{r}_{j})$ is
sometimes called the structure factor (of the simulation box).\\ 
The same procedure, with the same damping function, has been applied to
systems in two dimensions \cite{r5,r6,r34}. In particular for Coulomb
interaction, it gives 
\beq
\left\{\begin{array}{ll}
&\displaystyle E_{\bm{s}}^{(s)}=\frac{1}{2}\sum_{i=1}^{N}\sum_{j=1}^{N}\sum_{\bm{\nu}}'q_{i}q_{j}\frac{\mbox{erfc}(\alpha|\bm{s_{ij}}+\bm{L}_{\bm{\nu}}|)}{|\bm{s_{ij}}+\bm{L}_{\bm{\nu}}|}\\
&\\
&\displaystyle E_{\bm{\kappa}=\bm{0}}^{(l)}=-\frac{\sqrt{\pi}}{\alpha A}\mbox{\large(}\sum_{i=1}^{N}q_{i}\mbox{\large)}^{2}-\frac{\alpha}{\sqrt{\pi}}\sum_{i=1}^{N}q_{i}^{2}\\
&\\
&\displaystyle E_{\bm{\kappa}\neq\bm{0}}^{(l)}=-\frac{\pi}{2A}\sum_{\bm{\kappa}\neq\bm{0}}\sum_{i=1}^{N}\sum_{j=1}^{N}q_{i}q_{j}\frac{\mbox{erfc}(\kappa/2\alpha)}{\kappa}\exp(i\bm{\kappa}.\bm{s}_{ij})
\end{array}
\right.
\eeq
with 
\begin{center}
$\displaystyle \bm{L}_{\bm{\nu}}=\nu_{x}L_{x}\hat{\bm{e}}_{x}+\nu_{y}L_{y}\hat{\bm{e}}_{y}$
\end{center}
where $\nu_{x}$ and $\nu_{y}$ are relative integers.\\
$\bm{\kappa}$ are the lattice vectors in the bidimensional Fourier
space, $\bm{s}_{i}=(x_{i},y_{i})$ the position of the particles in the
two dimensional space and $\bm{s}_{ij}=\bm{s}_{i}-\bm{s}_{j}$. The
convergence parameter $\alpha$ is usually chosen such as the summations
that give $E_{\bm{r}}^{(s)}$ can be restricted to
$\bm{\nu}=\bm{0}$. This choice is often taken as $\alpha L\simeq 5-8$.\\   
For systems with a quasi-two dimensional geometry, as the bilayer Wigner
crystal described in section II, the Ewald method for three dimensional
systems, given by Eqs.(7), may be applied by taking a highly asymmetric
box [35,20,21(b)]. For systems with a slab geometry, this procedure is
certainly the most efficient [21(b),24]. A careful examination of the
summations may show that large errors, arising by a naive implementation
of the tridimensional Ewald method can be corrected analytically by
adding some corrections terms [20,21(b)]. Nevertheless, these
procedures may not easily be applied to systems that do not have the
specific symmetry of slab-like systems, as it is the case for
fluid-fluid, fluid-solid interfaces, monolayers, etc. For these reasons,
several methods have been developed to take into account the finite
extension of systems without any periodicity in the direction of the
finite extension.\\   
The method for real two dimensional systems, given by Eqs.(8), may of
course not be applied to quasi-two dimensional systems, except as a
limiting case when the length of the finite extension in the third
direction tends to zero ($h\rightarrow 0$).  

\subsection{The Ewald method for quasi-two dimensional systems.}

Following analytical computations done in refs.[10-15], one may derived an
Ewald method for three dimensional systems with periodicity in only two
directions.\\   
One of the most simple derivation of the Ewald quasi-2D method is the
analytical derivation done by Parry \cite{r10} in 1975; this derivation of
the Ewald quasi-2D method is also very useful to link this method to the
Ewald method for tridimensional systems. In the derivation done by
Parry, the Ewald-3D method is taken as a starting point. By letting the
periodicity of the periodic boundary condition in the direction where
the system has a finite extension tending to infinity, an analytical
formulation of the Ewald quasi-2D method is found. Other analytical
derivations, that do not use the Ewald-3D method as a starting point,
may also be found in refs.[11,13,15,35]. In particular, an elegant
method may be found in a recent work by Grzybowski, Gw\'o\'zd\'z, and
Br\'odka \cite{r15}; their derivation is also closely related to the rapidly
convergent representations for Green's functions for Laplace's equations
presented in a work by Linton \cite{r37}.\\  
We define our notations as following, 
\begin{center}
$\left\{\begin{array}{ll} 
\displaystyle \bm{r_{ij}}&\displaystyle =\bm{s}_{ij}+z_{ij}\bm{\hat{e}}_{z}\\
&\\
\displaystyle \bm{L}_{\bm{\nu}}&\displaystyle =\nu_{x}L_{x}\hat{\bm{e}}_{x}+\nu_{y}L_{y}\hat{\bm{e}}_{y}\\
&\\
\displaystyle \bm{L}_{\bm{n}}&\displaystyle=\bm{L}_{\bm{\nu}}+n_{z}L_{z}\hat{\bm{e}}_{z}\\
&\\
\displaystyle \bm{k}&\displaystyle =\bm{\kappa}+k_{z}\bm{\hat{e}}_{z}
\end{array}\right.$
\end{center}
where $k_{z}=2\pi/L_{z}$.\\ 
One finds by taking $L_{z}\rightarrow\infty$
\beq
E^{2D}=E_{\bm{r}}+E_{\bm{\kappa}=\bm{0}}+E_{\bm{\kappa}\neq\bm{0}}=E_{pp}^{(cv)}
\eeq
with the contributions given by \cite{r10}\\
\begin{picture}(450,50)
\put(10,20){$\displaystyle E_{\bm{r}}=\frac{1}{2}\sum_{i=1}^{N}\sum_{j=1}^{N}\sum'_{\bm{\nu}}q_{i}q_{j}\frac{\mbox{erfc}(\alpha|\bm{r_{ij}}+\bm{L}_{\bm{\nu}}|)}{|\bm{r_{ij}}+\bm{L}_{\bm{\nu}}|}$}
\put(400,20){(10.a)}
\end{picture}\\
\begin{picture}(450,50)
\put(10,20){$\displaystyle E_{\bm{\kappa}=\bm{0}}=-\frac{\pi}{A}\sum_{i=1}^{N}\sum_{j=1}^{N}q_{i}q_{j}\mbox{\Large (}\rst1\frac{\exp(-\alpha^{2}z_{ij}^{2})}{\alpha\sqrt{\pi}}+| z_{ij}|\mbox{erf}(\alpha| z_{ij}|)\mbox{\Large )}-\frac{\alpha}{\sqrt{\pi}}\sum_{i=1}^{N}q_{i}^2$}
\put(400,20){(10.b)}
\end{picture}\\
\begin{picture}(450,50)
\put(10,20){$\displaystyle E_{\bm{\kappa}\neq\bm{0}}=\frac{\pi}{2A}\sum_{\bm{\kappa}\neq\bm{0}}\sum_{i=1}^{N}\sum_{j=1}^{N}q_{i}q_{j}F(\kappa,\alpha;z_{ij})\frac{\exp(i\bm{\kappa}.\bm{s}_{ij})}{\kappa}$}
\put(400,20){(10.c)}
\end{picture}\\
\setcounter{equation}{10}
where
\beq
\displaystyle F(\kappa,\alpha;z_{ij})=\exp(\kappa z_{ij})\mbox{erfc}(\frac{\kappa}{2\alpha}+\alpha z_{ij})+\exp(-\kappa z_{ij})\mbox{erfc}(\frac{\kappa}{2\alpha}-\alpha z_{ij})
\eeq
and the contribution $\bm{\kappa}=\bm{0}$ arising from $k_{z}\neq 0$.\\
When applied to the bilayer Wigner crystal described in section II, one
may use the factorization as one particle summation in the long ranged
contributions given by Eq.(10.c). Thus, one may implement with the same
efficiency the Ewald quasi-2D method for this system as it is done in
the Ewald-3D method. This situation is very particular to the system
considered in the present work. For other quasi-2D systems, because the
complicated form of the function $F(\kappa,\alpha;z_{ij})$, one may not
in general find a factorization in one particle summation as in the
Ewald-3D method. For the bilayer Wigner crystal, $z_{ij}$ may take only
two values: if both particles are in the same layer, then $z_{ij}=0$ or
else $z_{ij}=h$.\\  
The intralayer contributions to the long ranged part of the electrostatic
energy ($z_{ij}=0$), for $\bm{\kappa}\neq \bm{0}$ are given by
\beq
\displaystyle E_{\bm{\kappa}\neq\bm{0}}^{\mbox{\footnotesize intra}}=\frac{\pi}{A}\sum_{\bm{\kappa}\neq\bm{0}}\frac{\mbox{erfc}(\kappa/2\alpha)}{\kappa}\mbox{\large(}\mbox{\large $\bracevert$}\sum_{i\in\L_{1}}q_{i}\exp(i\bm{\kappa}.\bm{s}_{i})\mbox{\large $\bracevert$}^{2}+\mbox{\large $\bracevert$}\sum_{i\in\L_{2}}q_{i}\exp(i\bm{\kappa}.\bm{s}_{i})\mbox{\large $\bracevert$}^{2}\mbox{\large)}
\eeq
and for $\bm{\kappa}=\bm{0}$ 
\beq
\displaystyle E_{\bm{\kappa}=\bm{0}}^{\mbox{\footnotesize intra}}=-\frac{\sqrt{\pi}}{\alpha A}\mbox{\large[}\mbox{\large(}\sum_{i\in\L_{1}}q_{i}\mbox{\large)}^{2}+\mbox{\large(}\sum_{i\in\L_{2}}q_{i}\mbox{\large)}^{2}\mbox{\large]}-\frac{\alpha}{\sqrt{\pi}}\mbox{\large(}\sum_{i\in\L_{1}}q_{i}^{2}+\sum_{i\in\L_{2}}q_{i}^{2}\mbox{\large)}
\eeq
Since the electroneutrality of the system is achieved with the surface
charge density of layers $\L_{1}$ and $\L_{2}$, contributions with
$(\sum_{i\in\L_{2}}q_{i})^{2}\neq 0$ have to be included in the
electrostatic energy (see appendix for more details).\\ 
The interlayer contributions to the long ranged part of the
electrostatic energy for $\bm{\kappa}\neq \bm{0}$ are given by 
\beq
\displaystyle E_{\bm{\kappa}\neq\bm{0}}^{\mbox{\footnotesize inter}}=\frac{\pi}{A}\sum_{\bm{\kappa}\neq\bm{0}}\frac{F(\kappa,\alpha;h)}{\kappa}\rst1 \Re\mbox{\large[}\mbox{\large(}\sum_{i\in\L_{1}}q_{i}\exp(i\bm{\kappa}.\bm{s}_{i})\mbox{\large)}\mbox{\large(}\sum_{j\in\L_{2}}q_{j}\exp(-i\bm{\kappa}.\bm{s}_{j})\mbox{\large)}\mbox{\large]}
\eeq
where $F(\kappa,\alpha;h)$ is given by Eq.(8) and $\Re[z]$ is the real
part of the complex number $z$. For $\bm{\kappa}=\bm{0}$, the interlayer
contribution given by Eq.(10.b) is 
\beq
\displaystyle E_{\bm{\kappa}=\bm{0}}^{\mbox{\footnotesize inter}}=-\frac{2\pi}{A}\mbox{\Large(}\rst1\frac{\exp(-\alpha^{2}h^{2})}{\alpha\sqrt{\pi}}+| h |\mbox{erf}(\alpha| h|)\mbox{\Large )}\mbox{\large(}\sum_{i\in\L_{1}}q_{i}\mbox{\large)}\mbox{\large(}\sum_{j\in\L_{2}}q_{j}\mbox{\large)}.
\eeq
The electrostatic energy of a configuration of the bilayer Wigner
crystal with the Ewald quasi-2D method is given by
\begin{center}
$\displaystyle E=E^{2D}+E_{W}=(E_{\bm{r}}^{\mbox{\footnotesize intra}}+E_{\bm{\kappa}=\bm{0}}^{\mbox{\footnotesize intra}}+E_{\bm{\kappa}\neq\bm{0}}^{\mbox{\footnotesize intra}})+(E_{\bm{r}}^{\mbox{\footnotesize inter}}+E_{\bm{\kappa}=\bm{0}}^{\mbox{\footnotesize inter}}+E_{\bm{\kappa}\neq\bm{0}}^{\mbox{\footnotesize inter}}+E_{W})$
\end{center}
with Eqs. (10.a), (12-15) and (A.14).\\
This expression of the electrostatic energy is an exact application of
the Ewald method to the bilayer system; all the computations with Monte
Carlo algorithm done with this method are considered in section IV as
reference points.\\
Generally, one may not implement with efficiency the Ewald quasi-2D
method, thus there are very few uses of this method. Of course, the
procedure described in this subsection has been used in the study of
the solid-phase diagram of the bilayer crystal \cite{r32} and also in other
numerical studies of electronic bilayer systems \cite{r38}. Most
applications of the Ewald quasi-2D method in computer simulations use
precalculated tables of potential energy, forces and second derivatives
on a three-dimensional grid and calculations are performed by
interpolation of the tables \cite{r39,r35,r20}.\\    
In ref.[40], Arnold and Holm have shown that the MMM2D method has a
computation effort that scales as $N^{5/3}(\log N)^{2}$. It is
interesting to note that for the bilayer Wigner system, it is possible
to implement the Ewald quasi-2D method with the same efficiency as the
Ewald 3D method, the computation effort scaling as $N^{3/2}$. Thus, for
this system, the Ewald quasi-2D method is slightly more efficient than
the MMM2D method.      

\subsection{The Hautman-Klein method.}

The computation of particle-particle interaction in the Hautman-Klein
method starts with the identity \cite{r16}
\beq
\displaystyle \frac{1}{r}=\frac{1}{r}-\sum_{n=0}^{M}a_{n}\frac{z^{2n}}{s^{2n+1}}+\sum_{n=0}^{M}a_{n}\frac{z^{2n}}{s^{2n+1}}
\eeq
with
\begin{center}
$\displaystyle a_{n}=\frac{(-1)^{n}(2n)!}{2^{2n}(n!)^{2}}$
\end{center}
By using damping functions $h_{n}(s;\alpha)$ on terms $1/s^{2n+1}$, one
separates the electrostatic energy of the system in a short ranged part
$E_{s}$ and long ranged part $E_{l}$, 
\begin{center}
$\displaystyle E^{HK}=E_{s}+E_{l}=E_{pp}^{(cv)}$
\end{center}
where
\beq
E_{s} \displaystyle = \frac{1}{2}\sum_{i=1}^{N}\sum_{j=1}^{N}q_{i}q_{j}\sum'_{\{\bm{\nu}\}}\mbox{\LARGE (}\frac{1}{r_{ij,\bm{\nu}}}-\sum_{n=0}^{M}a_{n}z_{ij}^{2n}\frac{h_{n}(s_{ij,\bm{\nu}};\alpha)}{s_{ij,\bm{\nu}}^{2n+1}}\mbox{\LARGE )}
\eeq
and
\beq
E_{l} \displaystyle = \frac{1}{2}\sum_{i=1}^{N}\sum_{j=1}^{N}q_{i}q_{j}\sum_{n=0}^{M}a_{n}z_{ij}^{2n}\mbox{\LARGE (}\sum'_{\{\bm{\nu}\}}\frac{h_{n}(s_{ij,\bm{\nu}};\alpha)}{s_{ij,\bm{\nu}}^{2n+1}}\mbox{\LARGE )}
\eeq
As in the Ewald method, the long ranged contributions are evaluated by
using a Fourier transform. In the original derivation by J. Hautman and
M.L. Klein the damping functions $h_{n}(s;\alpha)$ are chosen such as  
\beq
\left\{\begin{array}{ll}
\displaystyle h_0(s;\alpha)&\displaystyle =\mbox{erf}(\alpha s)\\
&\\
\displaystyle\frac{h_{n}(s;\alpha)}{s^{2n+1}}&\displaystyle=\frac{1}{a_{n}(2n)!}\nabla^{2n}\mbox{\large (}\frac{h_{0}(s;\alpha)}{s}\mbox{\large )}\\
\end{array}
\right..
\eeq
The derivation done by J. Hautman and M.L. Klein with the damping
functions given in Eqs.(19) allows to recover easily the analytical
derivation of the Ewald method for exact two dimensional systems given
in Eqs.(8) by letting $z_{ij}\rightarrow 0$. This is obtained obviously
from equations (17-19). Using the damping functions of Eqs.(19), we have     
\beq
E_{s} \displaystyle = \frac{1}{2}\sum_{i=1}^{N}\sum_{j\neq i}q_{i}q_{j}\mbox{\LARGE (}\frac{1}{r_{ij}}-\frac{\mbox{erf}(\alpha s_{ij})}{s_{ij}}-\sum_{n=1}^{M}\frac{1}{(2n)!}z_{ij}^{2n}\nabla^{2n}\mbox{\large (}\frac{\mbox{erf}(\alpha s_{ij})}{s_{ij}}\mbox{\large )}\mbox{\LARGE )}
\eeq
and 
\beq
\begin{array}{ll}
E_{l} &\displaystyle =\frac{\pi}{A}\sum_{i=1}^{N}\sum_{j=1}^{N}q_{i}q_{j}\sum_{n=0}^{M}\frac{1}{(2n)!}z_{ij}^{2n}\sum_{\bm{\kappa}\neq\bm{0}}\kappa^{2n-1}\mbox{erfc}(\kappa/2\alpha)e^{i\bm{\kappa}.\bm{s}_{ij}}\\
&\\
&\displaystyle -\frac{\alpha}{\sqrt{\pi}}\sum_{i=1}^{N}q_{i}^{2}-\frac{\sqrt{\pi}}{\alpha A}\mbox{\large (}\sum_{i=1}^{N}q_{i}\mbox{\large )}^2
\end{array}
\eeq
According Eqs.(16) for $n=1$ to $3$ we have  
\beq
\left\{\begin{array}{lll}
&h_1(s;\alpha)&\displaystyle=\mbox{erf}(\alpha s)-\frac{2\alpha s}{\sqrt{\pi}}e^{-\alpha^2 s^2}(1+2\alpha^2 s^2)\\
&&\\
&h_2(s;\alpha)&\displaystyle=\mbox{erf}(\alpha s)-\frac{2\alpha s}{\sqrt{\pi}}e^{-\alpha^2 s^2}(1+\frac{2}{3}\alpha^2 s^2-\frac{4}{9}\alpha^4 s^4+\frac{8}{9}\alpha^6 s^6)\\
&&\\
&h_3(s;\alpha)&\displaystyle=\mbox{erf}(\alpha s)-\frac{2\alpha s}{\sqrt{\pi}}e^{-\alpha^2 s^2}(1+\frac{2}{3}\alpha^2 s^2+\frac{4}{15}\alpha^4 s^4+\frac{8}{25}\alpha^6 s^6\\
&&\\
&&\displaystyle\t2\t2\t2\t2 -\frac{112}{225}\alpha^8 s^8+\frac{32}{225}\alpha^{10} s^{10}))
\end{array}
\right.
\eeq
The attractive feature of the method is that, by writing the
$z_{ij}^{2n}$ explicitly as polynomials in $z_{i}$ and $z_{j}$, the long
ranged contribution given by Eq.(21) can be cast into a sum of terms
each of which involving product of two functions having the general form
\beq
F_{p}(\bm{\kappa})=\sum_{i=1}^{N}q_{i}z_{i}^{p}\exp(i\bm{\kappa}.\bm{s}_{i})
\eeq
A careful examination of Eqs.(20-21) in one hand and Eqs.(10) in the
other hand, shows that Hautman-Klein method may not be considered
exactly as Taylor expansion of the Ewald quasi-2D method. In particular,
the complicated contribution $E_{\bm{\kappa}=\bm{0}}$ in Eq.(10.b) in
the Ewald quasi-2D method is roughly approximated in the Hautman-Klein
method by a constant. As a consequence, the contributions to intralayer
and interlayer energies do not have exactly the same constant terms.\\ 
For the bilayer Wigner crystal, the long ranged intralayer
contributions $E_{\bm{\kappa}\neq\bm{0},HK}^{\mbox{\footnotesize intra}}$ are exactly
given by Eq.(12) and with Eq.(21), one may expressed
$E_{\bm{\kappa}=\bm{0},HK}^{\mbox{\footnotesize intra}}$ as Eq.(13). Long ranged
interlayer contributions are given by 
\beq
\displaystyle E_{\bm{\kappa}\neq\bm{0},HK}^{\mbox{\footnotesize inter}}=\frac{\pi}{A}\sum_{\bm{\kappa}\neq\bm{0}}\frac{H(\kappa,\alpha;h)}{\kappa}\rst1\Re\mbox{\large[}\mbox{\large(}\sum_{i\in\L_{1}}q_{i}\exp(i\bm{\kappa}.\bm{s}_{i})\mbox{\large)}\mbox{\large(}\sum_{j\in\L_{2}}q_{j}\exp(-i\bm{\kappa}.\bm{s}_{j})\mbox{\large)}\mbox{\large]}
\eeq
with the same notations as in Eq.(14) and 
\beq
H(\kappa,\alpha;h)=\sum_{n=0}^{M}\frac{1}{(2n)!}(\kappa h)^{2n}\mbox{erfc}(\kappa/2\alpha)
\eeq
The contribution to interlayer energy for $\bm{\kappa}=\bm{0}$,
according to Eqs.(21) and (13) has to be written as 
\beq
\displaystyle E_{\bm{\kappa}=\bm{0},HK}^{\mbox{\footnotesize inter}}=-\frac{2\sqrt{\pi}}{\alpha A}\mbox{\large(}\sum_{i\in\L_{1}}q_{i}\mbox{\large)}\mbox{\large(}\sum_{j\in\L_{2}}q_{j}\mbox{\large)}.
\eeq
instead of the Taylor expansion of Eq.(15).\\
The electrostatic energy of a configuration of the bilayer Wigner
crystal with the Hautman-Klein method is given by
\begin{center}
$\begin{array}{ll}
\displaystyle  E=E^{HK}+E_{W}&\displaystyle =(E_{\bm{r},HK}^{\mbox{\footnotesize intra}}+E_{\bm{\kappa}=\bm{0},HK}^{\mbox{\footnotesize intra}}+E_{\bm{\kappa}\neq\bm{0},HK}^{\mbox{\footnotesize intra}})\\
&\\
&\displaystyle+(E_{\bm{r},HK}^{\mbox{\footnotesize inter}}+E_{\bm{\kappa}=\bm{0},HK}^{\mbox{\footnotesize inter}}+E_{\bm{\kappa}\neq\bm{0},HK}^{\mbox{\footnotesize inter}}+E_{W})\end{array}$
\end{center}
with Eqs.(20), (12), (13), (24), (26) and (A.14).\\
Because an expansion of $1/r$ for small $z$ is taken as a starting point
in Hautman-Klein method, this method is often considered as inaccurate
for rather thick systems. Despite its obvious inaccuracy, for slab-like
systems, as the system studied in ref.[25], the results found with the
Hautman-Klein method are in very good agreement with results obtained
with a purely electrostatic model and results obtained with the methods
of refs.[18,20].\\  
The Hautman-Klein method may also be implemented with almost the same
efficiency as Ewald methods for bulk-like systems, but without taking
any periodic images in the direction of the finite extension of the
system. Consequently, the Hautman-Klein method is specifically well
adapted to systems as fluid-fluid or fluid-solid interfaces, that do not
have a slab-like geometry. The Hautman-Klein method may also be
implemented to take into account electrostatic images at the interface
between two media with different dielectric constants. This method has
been used recently in a study of a mixed lipid/non-ionic surfactant
membrane \cite{r41}.  

\subsection{The Lekner method.}

The third procedure used in the present work is a method proposed by
J. Lekner \cite{r19}. This method may also be more or less easily
related to the original Ewald method.\\ 
When applied to quasi-one dimensional systems, Lekner summations method
is an eigenfunction expansion of the Green function of Laplace equation
with periodicity in only one direction (see for instance Eq.(3.10) in
ref.[37]). Following the derivations done in refs.[13,15,37], an Ewald
method for quasi-one dimensional systems may be derived \cite{r42}, the
Lekner summations for the quasi-one dimensional systems may be recovered
by letting the convergence parameter $\alpha$, defined in Eq.(4), tends
to infinity \cite{r37,r43}.\\      
For quasi-two dimensional systems, the relation between Lekner
summations and Ewald summations is sligthly more complicated. With the
derivation that permits in quasi-one dimensional systems to relate the
Lekner summations to an eigenfunction expansion of Green functions, one
is obtaining for quasi-two dimensional systems the so-called Nijboer-de
Wette representation \cite{r12,r13,r44}, the same method as the one used
recently in the method called MMM2D (see ref.[21(a)] and [45]); Lekner
summations are related to these methods by an asymmetric use of the
Poisson-Jacobi formula. As for quasi-one dimensional geometries, when
the convergence parameter $\alpha$ tends to infinity, one may relate
easily the Nijboer-de Wette representation (or the 'far formula' of the
MMM2D method \cite{r21}) to the Ewald method \cite{r37}.\\  
Following the original derivation done by J. Lekner \cite{r19} and
computations done by N. Gr$\o$nbech-Jensen and co-workers [46-49] and
by S. Marshall \cite{r50}, the electrostatic energy of a quasi-two
dimensional system can be computed as 
\beq
E^{LC}=\frac{1}{2}\sum_{i=1}^{N}\sum_{j=1,j\neq i}^{N}E_{ij}+\sum_{i=1}^{N}U_{i}^{\mbox{\footnotesize Self}}
\eeq 
The interaction energy $E_{ij}$ between pairs is approximated in a
numerical computation by one of the two formulas
\beq
\begin{array}{ll}
\displaystyle V_{ij}(n_{c},n_{K})&\displaystyle =4\frac{q_{i}q_{j}}{L_{y}}\sum_{m=1}^{n_{c}}\cos(2\pi\frac{y_{ij}}{L_{y}}m)\sum_{k=-n_{K}}^{n_{K}}K_{0}\mbox{\Large{[}}2\pi m\mbox{\large{[}}\mbox{\large{(}}\frac{L_{x}}{L_{y}}\mbox{\large{)}}^{2}\mbox{\large{(}}\frac{x_{ij}}{L_{x}}+k\mbox{\large{)}}^{2}+\mbox{\large{(}}\frac{z_{ij}}{L_{y}}\mbox{\large{)}}^{2}\mbox{\large{]}}^{\frac{1}{2}}\mbox{\Large{]}} \\
&\displaystyle \\
&\displaystyle -\frac{q_{i}q_{j}}{L_{y}}\ln\mbox{\large{[}}\cosh\mbox{\large{(}}2\pi\frac{z_{ij}}{L_{x}}\mbox{\large{)}}-\cos\mbox{\large{(}}2\pi\frac{x_{ij}}{L_{x}}\mbox{\large{)}}\mbox{\large{]}}-\frac{q_{i}q_{j}}{L_{y}}\ln 2
\end{array}
\eeq
or
\beq
\begin{array}{ll}
\displaystyle U_{ij}(n_{c},n_{K})&\displaystyle =4\frac{q_{i}q_{j}}{L_{x}}\sum_{n=1}^{n_{c}}\cos(2\pi\frac{x_{ij}}{L_{x}}n)\sum_{k=-n_{K}}^{n_{K}}K_{0}\mbox{\Large{[}}2\pi n\mbox{\large{[}}\mbox{\large{(}}\frac{L_{y}}{L_{x}}\mbox{\large{)}}^{2}\mbox{\large{(}}\frac{y_{ij}}{L_{y}}+k\mbox{\large{)}}^{2}+\mbox{\large{(}}\frac{z_{ij}}{L_{x}}\mbox{\large{)}}^{2}\mbox{\large{]}}^{\frac{1}{2}}\mbox{\Large{]}} \\
&\displaystyle\\
&\displaystyle -\frac{q_{i}q_{j}}{L_{x}}\ln\mbox{\large{[}}\cosh\mbox{\large{(}}2\pi\frac{z_{ij}}{L_{y}}\mbox{\large{)}}-\cos\mbox{\large{(}}2\pi\frac{y_{ij}}{L_{y}}\mbox{\large{)}}\mbox{\large{]}}-\frac{q_{i}q_{j}}{L_{x}}\ln 2
\end{array}
\eeq
The integers $n_{c}$ and $n_{K}$ are the truncation parameters that
would govern the accuracy of the finite summations \cite{r24,r37}. The main
technical problem that one would encounter in trying to implement
straightforwardly this procedure is in the convergence rate of one of
the two summations for some special configurations of the pair of
particles $i$, $j$ relatively to the geometry of the simulation box. For
these particular configurations, that are very frequent when few
billions of configurations are sampled, one of the two summations given
by Eqs.(28) and (29) is very slowly convergent and a lot of
contributions must be included to compute the correct value of the
energy. The origin of this slow convergence rate is in the behavior of
the modified Bessel function $K_{0}$ as its argument tends to zero (see
ref.[24]). To implement correctly the Lekner summations in a Monte Carlo
algorithm or in a Molecular dynamics simulation, one must use a
procedure to overcome this poor convergence rate or else some bias would
plague the simulation. To achieve this aim, several methods have been
proposed. A procedure frequently used consist in applying a cyclic
symmetry that is found in the original derivation of the
method \cite{r19}. In the following, we restrict ourself to this procedure
and the implementation of the Lekner summations according to this
procedure is called the Lekner-cyclic method. At the end of this
subsection we will review briefly other procedures proposed to overcome
the poor convergence rate of one of the summations $U_{ij}(n_{c},n_{K})$
or $V_{ij}(n_{c},n_{K})$.\\    
The cyclic symmetry consists in recognizing that as
$n_{c}\rightarrow\infty$ and $n_{K}\rightarrow\infty$, we have  
\beq
\lim_{n_{c}\rightarrow\infty,n_{K}\rightarrow\infty}V_{ij}(n_{c},n_{K})=\lim_{n_{c}\rightarrow\infty,n_{K}\rightarrow\infty}U_{ij}(n_{c},n_{K})=E_{ij}
\eeq  
For the particular configurations where one of the two summations is
slowly convergent, to compute $E_{ij}$ one may use $V_{ij}(n_{c},n_{K})$
or $U_{ij}(n_{c},n_{K})$ (refs.[19,47,50]). Since for given $n_{c}$ and
$n_{K}$, $V_{ij}(n_{c},n_{K})$ and $U_{ij}(n_{c},n_{K})$ may have very
different numerical values, the application of the cyclic symmetry may
introduce complicated bias \cite{r24}. In particular, if $n_{c}$ and $n_{K}$
are badly chosen, the value for $E_{ij}$ found in computing energy of
particle $i$ in the electrostatic potential of the particle $j$ may be
different of the value found for $E_{ji}$ when the energy is computed by
considering the particle $j$. An inequality as $E_{ij}\neq E_{ji}$
happens here because the configuration of the vector $\bm{s}_{ij}$,
relatively to the simulation box, is different from the configuration of
the vector $\bm{s}_{ji}$, especially when the minimum image convention
is used.\\   
The study done in ref.[24] is helpful to define a criterion on
$\bm{s}_{ij}$ to choose to compute $V_{ij}(n_{c},n_{K})$ rather than
$U_{ij}(n_{c},n_{K})$ or the converse.\\ 
For the bilayer Wigner crystal, the interaction energy $E_{ij}$ between
pair of particles is computed as
\beq
E_{ij}=\left\{\begin{array}{ll}
	V_{ij}(n_{c},n_{K})\t2\t2&\displaystyle\mbox{if}\t2\t2| x_{ij}| > | y_{ij}|\\
	&\\
	U_{ij}(n_{c},n_{K})\t2\t2&\displaystyle\mbox{if}\t2\t2| x_{ij}| < | y_{ij}|
	\end{array}
	\right.
\eeq
The criterion used here is quite simple because shapes of the layers are
squares of side $L$.\\
General formulas as Eqs.(28) and (29) for Lekner summations, when the
base of unit cell is not a square ($L_{x}\neq L_{y}$) or do not have a
rectangular shape, have been proposed in refs.[48,51-53]; for these
geometries a careful examination of the criterion given by Eq.(31)
should be done, especially when the base of the unit cell is not
rectangular.\\   
To compute, with the cyclic symmetry, the energy of a configuration made
of $N$ particles, the number of different contributions to evaluate is
of the order $n_{c}(2n_{K}+1)N^{2}$. As it was shown in our previous
work \cite{r24}, when the criterion given by Eq.(31) is used, for some
configurations of the vector $\bm{s}_{ij}$ the value computed by using
the most rapidly convergent of the two formulas (28) and (29) becomes
independent of the direction of the vector $\bm{s}_{ij}$ for rather high
values of $n_{c}$ (e.g. $n_{c}\simeq 40$ for a relative accuracy of
$10^{-5}-10^{-6}$) \cite{r24,r37}. On the other hand, for most of the
configurations of the vectors $\bm{s}_{ij}$, the convergence of
summations (28) and (29) are more rapids, thus a value as large as
$n_{c}\simeq 40$ is not needed in all computations done with the Lekner
summations. Then, to improve the efficiency of the method, $n_{c}$ is
not chosen independently of the argument of the Bessel function. Taking
into account the asymptotic behavior of the Bessel function $K_{0}(x)$ 
\begin{center}
$\displaystyle K_{0}(x)\simeq
\sqrt{\frac{\pi}{2x}}\exp(-x)\t2\t2\mbox{as}\t2\t2 x\rightarrow\infty $
\end{center}
truncation to $n_{K}=3$ or 4 is enough to achieve a good accuracy
($K_{0}(19.)\simeq 1.6\rst1 10^{-9}$). The value of $n_{c}$ used is
chosen such as
\beq
1\leq n_{c}(\bm{s}_{ij})\leq n_{c}^{max}
\eeq
where $n_{c}^{max}$ is a prefixed value and $n_{c}(\bm{s}_{ij})$ the
value of $n_{c}$ chosen in such a way that the output of the summation,
Eq.(28) or Eq.(29), would be independent of the vector $\bm{s}_{ij}$. In
the computations presented in the next section, $n_{c}(\bm{s}_{ij})$ is
chosen as the first integer such as  
\beq
\displaystyle 2\pi n_{c}(\bm{s}_{ij})\mbox{\large{[}}\mbox{\large{(}}\frac{y_{ij}}{L}+k\mbox{\large{)}}^{2}+\mbox{\large{(}}\frac{z_{ij}}{L}\mbox{\large{)}}^{2}\mbox{\large{]}}^{\frac{1}{2}}>19.
\eeq
If the integer found by using Eq.(33) is greater than $n_{c}^{max}$, then
one takes $n_{c}(\bm{s}_{ij})=n_{c}^{max}$.\\
For each value of the term between the brakets in Eq.(33), a value of
$n_{c}(\bm{s}_{ij})$ is taken from a precomputed table that takes into
account the prescription imposed by $n_{c}^{max}$. This evaluation of
$n_{c}(\bm{s}_{ij})$ is done for all pairs of particles. For the bilayer
Wigner crystal this procedure is used for both the intralayer and
interlayer contributions.\\ 
This procedure has been applied to all computations done with the
Lekner-cyclic method presented in section IV, in this section $n_{c}$
denotes the value of $n_{c}^{max}$ used in Eqs.(32) and (33).\\ 
For the bilayer Wigner crystal, the separation of the energy into
intralayer and interlayer contributions is done as in the Ewald quasi-2D
and Hautman-Klein methods as 
\beq
E^{LC}=E^{LC}_{\mbox{\footnotesize intra}}+E^{LC}_{\mbox{\footnotesize inter}}=E_{pp}^{(cv)}
\eeq
with
\beq
E^{LC}_{\mbox{\footnotesize intra}}=\frac{1}{2}\sum_{i\in\L_{1}}\sum_{j\in\L_{1};j\neq i}E_{ij}+\frac{1}{2}\sum_{i\in\L_{2}}\sum_{j\in\L_{2};j\neq i}E_{ij}+\sum_{i\in\L_{1}}U_{i}^{\mbox{\footnotesize Self}}+\sum_{i\in\L_{2}}U_{i}^{\mbox{\footnotesize Self}}
\eeq
where
\beq
\displaystyle U_{i}^{\mbox{\footnotesize Self}}=\frac{1}{2}\lim_{\bm{r}_{i}\rightarrow\bm{r}_{j}}\mbox{\large{(}}E_{ij}-\frac{q_{i}^{2}}{r_{ij}}\mbox{\large{)}}=\frac{q_{i}^{2}}{L}\mbox{\large{[}}4\sum_{m=1}^{\infty}\sum_{k=1}^{\infty}K_{0}\mbox{\large{(}}2\pi mk\mbox{\large{)}}+\gamma-\ln(4\pi)\mbox{\large{]}}
\eeq
and
\beq
E^{LC}_{\mbox{\footnotesize inter}}=\sum_{i\in\L_{1}}\sum_{j\in\L_{2}}E_{ij}
\eeq
For the bilayer Wigner crystal, we have 
\beq
\begin{array}{ll}
\displaystyle V_{ij}^{\mbox{\footnotesize intra}}(n_{c},n_{K})&\displaystyle=4\frac{q_{i}q_{j}}{L}\sum_{m=1}^{n_{c}}\cos(\frac{2\pi m}{L}y_{ij})\sum_{k=-n_{K}}^{n_{K}}K_{0}\mbox{\Large{[}}2\pi m| \frac{x_{ij}}{L}+k| \mbox{\Large{]}} \\
&\displaystyle \\
&\displaystyle -\frac{q_{i}q_{j}}{L}\ln\mbox{\large{[}}1-\cos\mbox{\large{(}}2\pi\frac{x_{ij}}{L}\mbox{\large{)}}\mbox{\large{]}}-\frac{q_{i}q_{j}}{L}\ln 2
\end{array}
\eeq
for intralayer contributions, and
\beq
\begin{array}{ll}
\displaystyle V_{ij}^{\mbox{\footnotesize inter}}(n_{c},n_{K})&\displaystyle =4\frac{q_{i}q_{j}}{L}\sum_{m=1}^{n_{c}}\cos(\frac{2\pi m}{L}y_{ij})\sum_{k=-n_{K}}^{n_{K}}K_{0}\mbox{\Large{[}}2\pi m\mbox{\large{[}}\mbox{\large{(}}\frac{x_{ij}}{L}+k\mbox{\large{)}}^{2}+\mbox{\large{(}}\frac{h}{L}\mbox{\large{)}}^{2}\mbox{\large{]}}^{\frac{1}{2}}\mbox{\Large{]}} \\
&\displaystyle \\
&\displaystyle -\frac{q_{i}q_{j}}{L}\ln\mbox{\large{[}}\cosh\mbox{\large{(}}2\pi\frac{h}{L}\mbox{\large{)}}-\cos\mbox{\large{(}}2\pi\frac{x_{ij}}{L}\mbox{\large{)}}\mbox{\large{]}}-\frac{q_{i}q_{j}}{L}\ln 2
\end{array}
\eeq
for interlayer contributions and similar relations for
$U_{ij}^{\mbox{\footnotesize intra}}$ and $U_{ij}^{\mbox{\footnotesize
inter}}$ (interchanging $x_{ij}$ with $y_{ij}$).\\  
The electrostatic energy of a configuration of the bilayer Wigner
crystal with the Lekner-cyclic method is given by
\beq
E=E^{LC}+E_{W}=E^{LC}_{\mbox{\footnotesize intra}}+(E^{LC}_{\mbox{\footnotesize inter}}+E_{W})
\eeq
with equations (35) to (39) and (A.14).\\
The Lekner summations method has been used in few studies using
molecular simulation algorithms.\\
The cyclic symmetry has been used recently in a Molecular Dynamics study
of thin water-acetonitrile films \cite{r54}: in this work an appendix that
describes the method used to implement the Lekner-cyclic method may be
found. In recent Monte Carlo simulations of adsorption of colloidal
particles on a charged surface \cite{r55}, the Lekner-cyclic method is also
used \cite{r56}, but technical details on the implementation are not
given in ref.[55].\\ 
To implement correctly the Lekner summations method, other procedures
have been proposed. In the present work, none of these procedures,
reviewed briefly below, have been neither implemented, nor compared to
the Lekner-cyclic or Ewald quasi-2D methods.\\  
In ref.[57], J.F. Harper proposed to use an integral representation of
summations when arguments of Bessel functions are small. This method
might be well adapted for quasi-one dimensional systems \cite{r42} where,
instead of having two different analytical relations, Eqs.(28) and (29),
one has only one summation. In these situations one cannot use the
cyclic symmetry.\\   
Another procedure is to derive an alternative expression which converges
fast as the argument of the Bessel functions tends to zero. Using the
Hurwitz zeta function, R. Sperb  had proposed such alternative
expressions \cite{r58}. The Sperb method has been used in a Molecular
Dynamics study of strongly charged polyelectrolyte brushes \cite{r57}; in
this work one would find an appendix that describes the procedures used
by authors. In ref.[43], A. Grzybowski and A. Br\'odka also used a
procedure close to the Sperb method for quasi-one dimensional
systems. The Lekner-Sperb method was also used by A.G. Moreira and
R.R. Netz in their studies of counterions distributions near charged
plates \cite{r60,r61}. In ref.[61], Moreira and Netz give a detailled and
very clear description of their implementation of Lekner summations. In
their implementation they use Sperb transformation with the following
procedure: if $\rho=\sqrt{(y+k)^{2}+z^{2}}\geq 1/3$, they use relations
similar to Eqs.(28) and (29) of the present work with truncation
parameters $(n_{c},n_{K})\simeq (11,2)$, if $\rho < 1/3$ Sperb formula
is used. As it will be shown in section IV, when Lekner-cyclic method is
used, truncation parameters as $(n_{c},n_{K})\simeq (10,3)$ do not give
accurate results. Nevertheless, one should not conclude that the
truncation parameters used by Moreira and Netz are too small to give
accurate results; in the core of their method they use Sperb formula
which main purpose is to reduce these truncation
parameters. Implementation done in ref.[61] and in the present work are
not exactly identicals.\\   
Another procedure has been developed by J.N. Newman based on
two-dimensional Chebyshev expansions and economized polynomials \cite{r62}.\\ 
There are few others studies that used Lekner summations on very
different, interesting and complicated systems [55,63-67], but details
on implementation are not given there. Although the cyclic symmetry is
certainly the procedure the mostly used for quasi-two dimensional
systems, details on the implementation are needed. Firstly, there are at
least three different methods to implement the Lekner summations
method. Secondly, as it will be shown in section IV, the results of a
computation may be depending on parameters and criterions used to
correct the slow convergence rate of one of the two summations given by
Eqs.(28) and (29). Moreover, the criterions used to implement the Lekner
cyclic method in Eqs.(31), (32) and (33) are depending on the
geometrical parameters of the simulation box or on parameters of the
studied system.     

\section{Results.} 

In this section, we present the results obtained with the three methods
of the previous section.\\
The three computer codes used in this work are very similar. The main
differences between the three programs are in the subroutines that
compute the electrostatic energy, all others parts of the programs are
the same. In particular, the files in which configurations are written
may be used in any of the three programs, the random numbers generator
used is the same and it is possible, with our codes, to start
computations for each of the three methods from exactly the same initial
condition and with exactly the same sequence of random numbers for the
trial moves of the particles and for the acceptation tests. This point
was necessary  not only to compare the compatibility of the methods, but
also to test the efficiency and the accuracy of the methods.\\ 
This section is made of two subsections. In subsection IV.A, using the
same initial configuration and the same random numbers sequence, we
compare the sampling of the phase space of the bilayer Wigner crystal
done with the three methods. In subsection IV.B, the results of
computations for the Runs defined in TABLE I are given. All the
computations were performed on a Silicon Graphics Origin 2000
(R10000/195 processors). 

\subsection{Accuracy of the sampling of the phase space.}

In the Monte Carlo Metropolis algorithm \cite{r1,r2,r33}, the phase
space of the system studied is sampled by an importance-weighted random
walk. From an initial configuration of the system, its phase space is 
sampled by a construction of configurations in such a way that
configurations are generated with a probability proportional to the
statistical weight of the configurations. A convenient way is to
generate a trial configuration from an old configuration of the system
and to accept or reject the trial configuration according to an
acceptance probability defined with the help of the detailed balance
condition. According to this procedure, if the trial configuration is
accepted, it becomes the new configuration of the system, or else, if
the trial configuration is rejected, the old configuration is taken as
the new configuration. The procedure is implemented by building another
trial configuration from the new configuration.\\ 
When the trial configuration is generated only from an old configuration
and when the acceptance probability may only be defined with the
knowledge of the old and trial configurations, the chain of
configurations generated by the Metropolis algorithm has all the
properties of a Markov chain. All ensembles of the statistical mechanics
can be sampled with the Metropolis algorithm \cite{r1}.\\  
To apply the Monte Carlo algorithm to the canonical ensemble, the
simplest procedure is to generate trial configurations from old
configurations by a random displacement of one particle of the
system. The displacement of the particle (the new configuration of the
system) is then accepted according to a probability given by 
\beq
\displaystyle \mbox{acc}(o\rightarrow n)=\mbox{min}(1,\exp(-\beta(E_{n}-E_{o}))
\eeq
where $E_{n}$ is the total energy of the trial (new) configuration and
$E_{o}$ the total energy of the old configuration. For these procedures,
the accuracy of the computation of the energy is closely related to the
accuracy of the sampling of the phase space of the system.\\
For the bilayer Wigner crystal, the energy difference used in the
acceptance probability of Eq.(41) is computed by using one of the three
methods of section III and in the following we set 
\beq 
\displaystyle \beta \Delta E_{pp}^{(cv)}=-\beta(E_{n}-E_{o})
\eeq
Since for the Lekner-cyclic method, configurations where the point ions
are on a regular two dimensional lattices are very penalizing for the
convergence of summations, these kinds of configurations were not taken
as initial configurations. To generate initial configurations on which
an examination of the influence of the truncation parameters $n_{c}$ and
$n_{K}$ could be done, we have chosen to build initial configurations as
follows. $N_{0}$ point ions are arranged on a square lattice in each
layers; this configuration is taken as the initial configuration for the
code using the Ewald quasi-2D method. From this initial configuration,
100 MC-cycles were performed with the Ewald quasi-2D method. After these
100 MC-cycles ($200N_{0}$ trial moves), the point ions are no longer
arranged in a perfect regular lattice, thus this configuration may be
taken as an initial configuration for all the three methods. This
procedure presents also the advantage of allowing to start computations
from a well defined configuration that we may reproduce easily, if the
same sequence of random numbers is taken.\\  
If codes are written such as the random numbers are generated following
the same sequence and used for the same purpose in each of the three
codes, then, if the three methods give exactly the same energy, the
trajectory in the phase space obtained by the Markov chain of
configurations would be exactly the same in the three computations.\\
The trajectory followed by the system with the Ewald quasi-2D method is
considered as the exact trajectory; the trajectories obtained with the
Hautman-Klein and Lekner-cyclic methods are compared to this
trajectory. In TABLE II, we give some elements to appreciate the
accuracy of the sampling of the phase space by the Hautman-Klein and
Lekner-cyclic methods relatively to the sampling obtained with the Ewald
quasi-2D method for initial configuration that corresponds to Runs a
(cf. TABLE I). The energy difference $\beta \Delta E_{pp}^{(cv)}$ for
the first trial move of the first MC-cycle from the same initial
condition with the same sequence of random numbers is given for
different truncation parameters for the Hautman-Klein and Lekner-cyclic
methods. This first move is particularly interesting, whatever it is
accepted or rejected, the move is identical in the three methods and the
position of all the other particles are also the same in the three
methods. The values of $\beta \Delta E_{pp}^{(cv)}$ show that the
truncation parameter $n_{c}$ in the Lekner-cyclic method governs the
accuracy of the computations, while $n_{K}$ seems to have less incidence
on accuracy. In agreement with our previous work \cite{r24}, to reach a
relative accuracy of $10^{-4}-10^{-5}$, one needs to choose $n_{c}\simeq
40$, while $n_{K}=2$ or 3 seems to be sufficient. A careful examination
of other energy differences shows that, for few particular moves, one needs
to take $n_{K}=3$ to avoid a lost of accuracy; nevertheless these events
are rare and in almost all cases, $n_{K}=2$ might be sufficient. On the
other hand, $n_{c}<25$ always leads to important differences in almost
all energy differences.\\    
Because energy differences in the three methods are not exactly the same,
the trajectory in the phase space followed by the system are not exactly
the same. In TABLE II, we give also the number of trial moves accepted
after the first MC-cycle (t=1) and after the first hundred MC-cycles
(t=100) from the same initial condition. In a deterministic point of
view, the trajectory in the phase space, using the Metropolis algorithm,
has to be independing of the method used to compute the energy. Of
course, in almost all molecular simulations, such deterministic point of
view is unreachable, at least because of round-off errors done by the
machines on floating points \cite{r68}. In our study, as soon as a random
trial move in the computation using the Ewald quasi-2D method is
accepted (or rejected), while it is rejected (or accepted) in
computation using Hautman-Klein of Lekner-cyclic methods, the
trajectories are no longer the same. Since in all situations examinated,
one may not exactly follow the same trajectory with the Hautman-Klein
and Lekner-cyclic methods as with the Ewald quasi-2D method, we should
find a criterion to estimate if these methods are enough accurate to
give reliable statistical averages.\\    
To answer this question, one has to come back to the principles of the
Monte Carlo sampling. To obtain reliable thermodynamical averages by
making a Monte Carlo sampling, one has to generate configurations with a
probability that would be proportional, with a rather good accuracy, to
the statistical weight of the configurations. Thus, it is not necessary
that the trajectories followed by the system will be exactly the same,
even if the initial configurations and the sequence of random numbers are
the same. The important point is to generate configurations with a
probability proportional to the statistical weight of the configurations;
if this is not achieved, the thermodynamical averages are biased in a more
or less complicated way. To estimate the bias from the 'exact' results
given by the Ewald quasi-2D method, we have chosen to follow the
instantaneous energy per particle. In Figure 2 (a-d), we have plotted
for the first 500 MC-cycles the instataneous energy per particle
obtained with the Ewald quasi-2D on one hand, and with others methods on
another hand, for the same initial configuration and with the same
sequence of random numbers. The intantaneous energy of the initial
configuration is given at $t=0$.\\  
In Figure 2.a, the energies using Hautman-Klein method ($M=3$) and Ewald
quasi-2d are shown. An examination of the first 50 MC-cycles shows that
both trajectories are exactly the same for these first moves. After, it
happens that a trial move is accepted in one of the two methods, while
it is rejected in the other. Trajectories separate at this
point. Nevertheless, fluctuations of this quantity are still in good
agreement in the two methods. This is an indication, but not a
demonstration, that both methods would give same statistical averages
and thus it seems that there are no bias, relatively to the Ewald
quasi-2D method, in using Hautman-Klein method for the geometrical
parameters corresponding to Runs a.\\  
The situation in Figure 2.b, where the same quantities are plotted for
Ewald quasi-2D method and Lekner-cyclic method with
$(n_{c},n_{K})=(10,3)$, is different. Trajectories separate in the first
trial moves of the first MC-cycle (see also TABLE II). For the first 50
MC-cycles, fluctuations of the instantaneous energy seem to be similar,
but after 100 MC-cycles the trajectories clearly separate, much more
than two magnitudes of the fluctuation. In this case, as it will be
shown in the next subsection, the statistical averages are different.\\  
Since for $n_{c}=10$ and $n_{K}=3$ as truncation parameters in the
Lekner-cyclic method the accuracy is not correct, we made the same
numerical analysis for higher value of these parameters. This is shown
on Figure 2.c and 2.d. In Figure 2.c the truncation parameters are
$n_{c}=25$ and $n_{K}=3$; the trajectories do not separate for the few
first trial moves but only after the few first MC-cycles, despite the
fact that the relative accuracy on the energy difference is only
$10^{-2}$ compared to the energy difference found for the Ewald quasi-2D
method. For the 500 MC-cycles shown in this Figure, with $n_{c}=25$ and
$n_{K}=3$, trajectories stay in good agreement. If one increases the
value of $n_{c}$, as it is done in Figure 2.d where $n_{c}=50$, one
would find a better agreement between Ewald quasi-2D and Lekner-cyclic
methods. For $n_{c}=50$ and $n_{K}=4$, the separation occurs only after
60 MC-cycles and after the trajectories stay very close. For both cases,
$(n_{c},n_{K})=(25,3)$ and $(n_{c},n_{K})=(50,4)$ one may hope to obtain
the same statistical averages than in the Ewald quasi-2D method. Of
course a dramatic separation, as for $(n_{c},n_{K})=(10,3)$, may still
happen for both $(n_{c},n_{K})=(25,3)$ and $(n_{c},n_{K})=(50,3)$ after
more MC-cycles, this had not happened for the first 10000 MC-cycles
examined.\\   
 
\subsection{Energy and pair distribution functions.}

To estimate statistical averages, we have mainly considered two
differents cases. A case where the separation between the two layers is
small ($h=1.0$) for which the coupling between the layers is important:
this corresponds to Runs a and b. We have also considered another case
where the separation between layers is large enough ($h=4.0$) so that
both layers are almost independent \cite{r32}, Runs c to f.\\       
Since for $h=1.0$, the coupling between the layers is important, this
has an incidence on the interlayer pair distribution functions
$g_{12}$. This separation between the layers allows us to estimate the
agreement between methods for small $z$. On the other hand, because for
$h=4.0$ the coupling between the two layers is very small, one would
find for almost all $s$, $g_{12}(s)\simeq 1$. One may estimate on these
computations with $h=4.0$ the influence of the periodic boundary
conditions by varying $N$.\\   
In TABLE III, the results for Runs a and b are given, $U/N$ is the
average energy per particle where $U$ is computed in the three methods
as 
\beq
U=<E>
\eeq
$E_{inter}/N$ and $E_{intra}/N$ are respectively the interlayer and
intralayer average energies per particle computed according section III;
and $\sigma_{U}$ is an estimation of the statistical fluctuation of the
total energy computed as   
\beq
\sigma_{U}^{2}=<E^{2}>-<E>^{2}
\eeq
On Figures 3 to 6, we give the intralayer $g_{11}(s)$ and interlayer
$g_{12}(s)$ pair distribution functions obtained with the Ewald quasi-2D
and other methods for Runs a and b.\\   
The value of $\sigma_{U}$ given in TABLE III can be used to give an
estimation of the accuracy needed on the average energy. The fluctuation
$\sigma_{U}$ is the average of the fluctuations observed on
Figures 2(a-d). The results given on TABLE III show that the Ewald quasi-2D
and Hautman-Klein methods in one hand and the Lekner-cyclic method with
$(n_{c},n_{K})=(25,3)$ on the other hand are in a very good agreement
for an estimation of the energy for Runs a and b. Using the results of TABLE
III, one may check that the difference between the average energy
obtained using Ewald quasi-2D method and Hautman-Klein method or
Lekner-cyclic method with $(n_{c},n_{K})=(25,3)$ is lesser than 2
$\sigma_{U}$. For Runs a, one has $\mid U_{Ewald}-U_{HK}\mid \simeq
0.2\rst1\sigma_{U}$ for Hautman-Klein method and $\mid
U_{Ewald}-U_{Lekner}\mid \simeq 0.4\rst1\sigma_{U}$ for Lekner-cyclic
method with $(n_{c},n_{K})=(25,3)$, while for Runs b, one has $\mid
U_{Ewald}-U_{HK}\mid \simeq 0.4\rst1\sigma_{U}$ for Hautman-Klein method
and $\mid U_{Ewald}-U_{Lekner}\mid \simeq 1.7\rst1\sigma_{U}$ for
Lekner-cyclic method with $(n_{c},n_{K})=(25,3)$. The lost of accuracy
in the Lekner-cyclic method with $(n_{c},n_{K})=(25,3)$ for Runs b may
be well understood by observing that as $L$ increases the argument of
the Bessel functions in Eqs.(38) and (39) decreases, thus one has to
increase $n_{c}$ to achieve a better accuracy. As shown on Figures 3 and
5 the pair distribution functions are also in good agreement, but with a
small lost of accuracy for Runs b in both Hautman-Klein method and
Lekner-cyclic method with $(n_{c},n_{K})=(25,3)$ compared to the Ewald
quasi-2D method. With $n_c$ raised to 40, one obtains a very good
agreement between the Ewald quasi-2D and Lekner-cyclic method for Runs a
and b, as it is shown on Figures 6.\\          
With the same criterions, it appears that the results for the Lekner
cyclic method with $(n_{c},n_{K})=(10,3)$ are not in a good agreement
with the Ewald quasi-2D method neither for Runs a, nor for Runs b. For
Runs a, one has $\mid U_{Ewald}-U_{Lekner}\mid \simeq
4.2\rst1\sigma_{U}$ (see also Fig.2(b)) and for Runs b $\mid
U_{Ewald}-U_{Lekner}\mid \simeq 5.6\rst1\sigma_{U}$. Then, as it was
mentioned in the previous subsection, the truncation parameters
$(n_{c},n_{K})=(10,3)$ are badly chosen to give reliable statistical
averages in this implementation of the Lekner-cyclic method. The bias
induced by a bad choice of the convergence parameters in Lekner
summations is exhibited on Figs.4(a-b). The pair distribution functions,
computed with the Lekner-cyclic method $(n_{c},n_{K})=(10,3)$ for Runs
a, differ dramatically from the pair distribution functions computed
with the Ewald quasi-2D method. As it is shown on Figs.5 (a-b), when the
value of the truncation parameter $n_{c}$ is raised to 25, pair
distribution functions $g_{11}$ and $g_{12}$ computed with the
Lekner-cyclic method are exactly the same as those computed with the
Ewald quasi-2D method. For Runs b, the observations done on Figs.4(c-d)
and Figs.5(c-d) are not as clear as it was for Runs a; nevertheless one
may not conclude that for the thermodynamical point corresponding to
Runs b, truncation parameters $(n_{c},n_{K})=(10,3)$ are enough to avoid
bias, since the average energy is very badly estimated. This point could
be misleading when testing the implementation of the Lekner method in a
computer code where no reference points, as those provided by the Ewald
quasi-2D method, can be found. As mentioned in TABLE I, for Runs b the
preferential structure of the layers is a disordered fluid-like
structure, thus there is less configurations of pairs of particles with
$x_{ij}\simeq 0$ or $y_{ij}\simeq 0$ compared to a crystal-like ordered
structure. As a consequence, summations given by Eqs.(38) and (39)
converge rapidly for most of pairs of particles and the bias induced by
the poor convergence rate of summations for few pairs of particles is
hidden. This situation becomes very dangerous when a fluid-solid
equilibrium is studied with these parameters. We must outlined also that
even if the average energy is badly evaluated with truncation parameter
$n_{c}=10$, the value found with this parameter is not very different
from the correct value given by the Ewald quasi-2D method, because of
the oscillatory behavior of summations given by Eqs.(38) and (39) (see
reference [24]). This point could also be misleading when tests on
implementation of the Lekner method are done.\\           
For $h=4.0$ (Runs c to f), the statistical averages are given on TABLE
IV with same notations. All these computations are done for the same
thermodynamical point but with different values of the number of
particles ranging from $N=128$ to $968$. According to the results of
TABLE IV, the Ewald quasi-2D and Hautman-Klein methods are in a good
agreement despite the large separation of layers ($h=4.0$) and except
for Runs c with N=128. For Runs d, e' and f, one has $\mid
U_{Ewald}-U_{HK}\mid\leq 0.7\rst1\sigma_{U}$. On the other hand, for
Runs c, where $N=128$ we have $\mid U_{Ewald}-U_{HK}\mid\simeq
3.5\rst1\sigma_{U}$ and the value found for $E_{inter}$ is not correct
with the Hautman-Klein method. This is illustrated on Figs.7 (a-b)
where it is shown that the pair distribution functions computed with
Ewald quasi-2D and Hautman-Klein methods are not in agreement. By using
the Hautman-Klein method with the geometrical parameters of Runs c, one
finds a coupling between both layers, while with the Ewald quasi-2D
method no coupling appears. This lost of accuracy in the Hautman-Klein
method for Runs c is introduced by the expansion of $1/r$ for small $z$,
one has $h/L\simeq 0.3$ for $N=128$, while for $N\geq 338$, we have $h/L
< 0.2$, but with the same density of point ions. Figures 7 and TABLE IV
show that to increase the accuracy of the Hautman-Klein method one may
increase $N$.\\ 
The finite-size effect is also outlined on Figures 7 for the Ewald
quasi-2D method. The pair distribution function $g_{11}$ is sensitive to
the finite size of the simulation box, this is illustrated in Fig.7(a)
up to Fig.7(g) where $N$ is ranging from 128 to 968.\\ 
Comparison between Ewald quasi-2D and the Lekner cyclic methods are done
in TABLE IV and in Figures 8 and 9, for Runs c to f. As for Runs a, the
Lekner-cyclic method with $(n_{c},n_{K})=(25,3)$ is in very good
agreement with the Ewald quasi-2D method. For Runs c to f, one has 
$\mid U_{Ewald}-U_{Lekner}\mid < 1.5\rst1\sigma_{U}$ for $N\geq 338$
and $\mid U_{Ewald}-U_{Lekner}\mid\simeq 1.8\rst1\sigma_{U}$ for
$N=128$. For $(n_{c},n_{K})=(10,3)$, TABLE IV shows that as $N$ is
increased, the Lekner-cyclic method lost its accuracy. For Runs c with
$(n_{c},n_{K})=(10,3)$, one has $\mid U_{Ewald}-U_{Lekner}\mid\simeq
1.5\rst1\sigma_{U}$ and for Runs f one finds  $\mid
U_{Ewald}-U_{Lekner}\mid\simeq 12\rst1\sigma_{U}$. This point is well
illustrated on the representation of $g_{11}$ given on Figures 8. As $N$
is increased from 128 to 968, the structure functions computed with the
Lekner-cyclic method become poorly evaluated. With
$(n_{c},n_{K})=(10,3)$, for $N=128$, $g_{11}$ and $g_{12}$ are in very
good agreement with the functions computed with the Ewald quasi-2D
method for $N=128$ (Figs.8 (a-b)), but as $N$ is increased, this
accuracy is lost. This lost of accuracy is similar to the one observed
in Runs b for the Lekner-cyclic method with $(n_{c},n_{K})=(25,3)$ since
as $N$ is increased, $L$ is increased and thus the argument of the
Bessel functions decreased. This point may also introduce bias when an
implementation of the Lekner method is done. Before making extensive
computations, one frequently checks the accuracy of the computation on
systems having very few particles. For systems with long ranged
interactions, a good estimation of some kind of Mandelung constants is
sometimes taken as a test of the method \cite{r19,r69}. Thus, with this
kind of tests, one may conclude using systems with very few particles
that a small value of $n_{c}$ gives accurate results, but this accuracy
would be lost as the number of particle is increased. Tests on systems
with a small numbers of particles (i.e. $N\leq 150$) are not sufficient
to check the accuracy of the implementation of the Lekner-cyclic
method. As it is shown on Figure 9, if one raises $n_c$ to 25, the
Lekner-cyclic and the Ewald quasi-2D methods are in very good agreement
for Runs c to f, both for the average energy and pair distribution
functions.\\   
On TABLES III and IV, we give also a crude estimation of the average
CPU-time in seconds per MC-cycle. These estimations are depending on
machines and codes. Improvement of the efficiency may certainly be
achieved; nevertheless these CPU-times may be intercompared to give an
estimation of the relative efficiency of the methods. The average
CPU-time per MC-cycle found for the Ewald quasi-2D method in TABLE IV
may not be connsidered as a representative efficiency of the method,
since for the particular system considered in this work the Ewald
quasi-2D is as efficient as an Ewald-3D method (see subsection
III.A). TABLE IV shows that the Lekner-cyclic method may become
particularly expensive as $N$ is increased (see also ref.[61]). With
Eq.(33), one may estimate the value of $n_{c}^{max}$ to achieve enough
accuracy with the Lekner-cyclic method as $N$ is increased, one roughly 
has   
\beq
\displaystyle n_{c}^{max}\sim \frac{19 L}{2\pi h}\sim N^{\frac{1}{2}}          
\eeq
Thus, the computing time with the implementation proposed in subsection
III.C scales like
\beq
\displaystyle \delta t \sim N^{\alpha}\t2\t2\t2\mbox{with}\t2\t2\t2 2\leq\alpha\leq\frac{5}{2}
\eeq

\section{Discussion} 
The study done in this paper gives a comparison between Lekner, Ewald
quasi-2D and Hautmann-Klein methods when implemented in a Monte Carlo
simulation. In particular the influence of the methods on
thermodynamical and structural quantities is outlined on a very simple
model on which the Ewald quasi-2D method may be  efficiently
implemented. Thus, the Ewald quasi-2D method and the bilayer Wigner
system may serve as reference tools to gain on physical quantities of
general interest, concretely (i.e. when implemented in Monte Carlo
Metropolis or in Molecular Dynamics simulations) the accuracy and
efficiency of all other methods proposed to handle quasi-bidimensional
systems.\\      
One of the main results of this paper is to demonstrate that the Ewald
quasi-2D method and the Lekner summation technique are in complete
agreement. In refs.[51,67], authors have compared the Lekner
method with the Hautman-Klein method; the outputs of the computations
presented in this paper agree with the results of these authors.\\ 
According the computations done here, we may not conclude that the
Lekner summations are an efficient alternative to the Ewald quasi-2D
method, at least for the implementation that uses the cyclic symmetry
(Lekner-cyclic method).\\
The procedure to implement the Lekner-cyclic method, described in the
subsection III.C of the present paper, is successful to use correctly
the Lekner summations in a Monte-Carlo Metropolis algorithm. This
procedure was based on a previous work where a study of the convergence
of the Lekner summations is done \cite{r24}. The truncation parameter
$n_{c}$ defined in ref.[24] and in Eqs.(28) and (29) governs the
accuracy of the sampling of the phase space of a system when Lekner
summations are used in Monte-Carlo Metropolis algorithms. Some severe
bias may be induced when the method is straigforwardly
implemented. These bias are generated by the diverging behavior of the
Bessel functions as its argument tends to zero. We may not recommend to
use a precise value of the truncation parameter $n_c$ to always
implement the Lekner method with accuracy, because the value of $n_c$
that must be used is depending on the geometrical parameters of systems
studied. Thus, for each system a convergence study is to make before
performing intensive computations.\\      
As explained in subsection IV.B, some tests may be misleading to check
the accuracy of the implementation. One of the most difficult behavior
to handle is the fact that the needed value of $n_{c}$ to achieve good
accuracy is depending on the number of particles, because the spatial
periodicity appears explicitly in the argument of the Bessel function
(see also Figures 8). The finite size effects in Lekner summations are
much more difficult to handle, than in Ewald methods; especially because
of the asymmetry of Eqs.(28) and (29).\\ 
In Ewald methods, it is required that energies are not depending on the
convergence parameter $\alpha$. The same requirement is to achieve with
the truncation parameter $n_{c}$ in Lekner methods.\\ 
The  proposed procedure in this present work to implement the Lekner
method in molecular simulations is accurate, but unfortunatly this
procedure is not very efficient. Others procedures have been proposed to
correctly implement the Lekner summations \cite{r55,r56,r60}. These
procedures have not been considered in the present work. Some of these
procedures may improved the efficiency of the method.\\     
The bilayer Wigner crystal used as a reference model in this work is
very convenient, since with the Ewald quasi-2D method exact results may
be obtained. Thus, this model and the results of TABLE III and TABLE IV
may be used as references to test the accuracy of other methods for
quasi-two dimensional systems.\\  
Lekner summations have also been used to compute Mandelung
constants \cite{r70}. In these computations, the Lekner summations are
useful and may be considered as an alternative to Ewald methods, mainly
because the spatial periodicities that are used in these computations
are much smaller than the spatial periodicities used in molecular
simulations.\\ 

\begin{center}
\large{\bf ACKNOWLEDGEMENTS}
\end{center}
We acknowledge the computation facilities of the $Centre$ $de$ $Ressources$
$Informatiques$ $(CRI)$ $de$ $l'Universit\acute{e}$ $de$
$Paris-Sud$. All the compuations done in this work were performed on a
Silicon Graphics Origin 2000 (R10000/195 processors).\\
I am very grateful to Professor J. Lekner for his sending of his work on
"$Energetics$ $of$ $hydrogen$ $ordering$ $in$ $ice$", Physica
B$\bm{252}$, 149 (1998) and to Dr. V. Huet for her help in the
preparation of the manuscript.   

\newpage
\begin{center}
\large{\bf Appendix : Electroneutrality and computation of the
interaction between two charged surfaces and ions.}
\end{center}
\renewcommand{\theequation}{A.\arabic{equation}}
\setcounter{equation}{0}
For the system described in Section II of the present work and more
generally for a system made of $N$ charged particles of charge $q$ and
two parallel surfaces ($\L_{1}\cup \L_{2}$) separated by a distance $h$
with a surface charge density $\sigma(\bm{r})$, electroneutrality is
given by  
\beq
Nq+\int\int_{\L_{1}\cup \L_{2}}d\bm{r}\sigma(\bm{r})=0
\eeq
Using conventional notations for the periodic boundary conditions, the
electrostatic energy of this system is given by
\beq
\begin{array}{lll}
\displaystyle E&\displaystyle=\frac{q^{2}}{2}\sum_{i=1}^{N}\sum_{j=1}^{N}\mbox{\LARGE{(}}\sum_{\bm{n}}'\frac{1}{\mid
\bm{r}_{ij}+\bm{n}\mid}\mbox{\LARGE{)}}&\displaystyle -q\sum_{i=1}^{N}\int_{\L_{1}\cup \L_{2}}d\bm{r}\sum_{\bm{n}}\frac{\sigma(\bm{r})}{\mid \bm{r}_{i}-\bm{r}+\bm{n}\mid}\\
&&\\
&&\displaystyle+\frac{1}{2}\int_{\L_{1}\cup \L_{2}}d\bm{r}d\bm{r}'\sum_{\bm{n}}'\frac{\sigma(\bm{r})\sigma(\bm{r}')}{\mid \bm{r}'-\bm{r}+\bm{n}\mid}\\
&&\\
&\displaystyle = E_{pp}+E_{pS}+E_{SS}
\end{array}
\eeq
The three contributions to the electrostatic energy are: (a) $E_{pp}$
the particle-particle interaction energy evaluated in this work by
using one of the three methods (i.e. Ewald quasi-2D, Hautman-Klein or
Lekner-cyclic methods) where self energies are included in $E_{pp}$, (b)
$E_{pS}$ is the particle-surface energy and (c) $E_{SS}$ the
surface-surface energy, including self energy of both surfaces.\\
In the following, we consider only uniform surface charge density and
using Eq.(A.1), one has
\beq
\sigma=-\frac{Nq}{2L^{2}}
\eeq
With these notations, the particle-surface energy is given by 
\beq
E_{pS}=-\frac{Nq^{2}}{2L^{2}}\sum_{i=1}^{N}\int_{\L_{1}\cup \L_{2}}d\bm{r}\sum_{\bm{n}}\frac{1}{\mid\bm{r}_{i}-\bm{r}+\bm{n}\mid}
\eeq
Following the derivation done by Grzybowski and co-workers in ref.[15]
and using the periodic boundary conditions, we have 
\beq
E_{pS}(\epsilon)=-\frac{Nq^{2}}{2 L^{2}}\sqrt{\pi}\sum_{i=1}^{N}\mbox{\LARGE{(}}f_{\epsilon}(\mid z_{i}-\frac{h}{2}\mid)+f_{\epsilon}(\mid z_{i}+\frac{h}{2}\mid)\mbox{\LARGE{)}}
\eeq
where
\beq
f_{\epsilon}(\mid Z\mid)=\int_{\epsilon}^{\infty}\frac{dt}{t^{\frac{3}{2}}}e^{-Z^{2}t}=2\sqrt{\pi}\mbox{\LARGE{[}}\frac{e^{-Z^2\epsilon}}{\sqrt{\epsilon\pi}}+\mid Z\mid\mbox{erf}(\sqrt{\epsilon}\mid Z\mid)\mbox{\LARGE{]}}-2\sqrt{\pi}\mid Z\mid
\eeq
The prescription $\epsilon$ has been introduced to outline the diverging
behavior of the integral in the second member of Eq.(A.6). As
$\epsilon\rightarrow 0$, we have 
\beq
f_{\epsilon}(\mid Z\mid)=\frac{2}{\sqrt{\epsilon}}-2\sqrt{\pi}\mid Z\mid+2\sqrt{\epsilon}\mid Z\mid^{2}+o(\epsilon^{\frac{3}{2}})
\eeq
As extensively commented in the literature, because of the
electroneutrality of the system, the diverging behavior as
$\epsilon\rightarrow 0$, is cancelled by adding all contributions to the
electrostatic energy (including self energy contributions). For systems
such as all particles are confined between charged surfaces
(i.e. $-h/2\leq z_{i}\leq h/2$), as the system studied in this work and
the system studied in ref.[25], we have
\beq
f_{\epsilon}(\mid z_{i}-\frac{h}{2}\mid)+f_{\epsilon}(\mid z_{i}+\frac{h}{2}\mid)=\frac{4}{\sqrt{\epsilon}}-\sqrt{\pi}h+2\sqrt{\epsilon}(\mid z_{i}-\frac{h}{2}\mid^{2}+\mid z_{i}+\frac{h}{2}\mid^{2})+o(\epsilon^{\frac{3}{2}})
\eeq
Following the same steps, the surface-surface contribution is 
\beq
E_{SS}(\epsilon)=\frac{N^{2}q^{2}}{2L^{2}}\frac{\sqrt{\pi}}{\sqrt{\epsilon}}+\frac{N^{2}q^{2}}{4L^{2}}\sqrt{\pi}f_{\epsilon}(h)
\eeq
When contributions $E_{SS}(\epsilon)$ and $E_{pS}(\epsilon)$ are added,
the diverging behavior is not yet cancelled by using electroneutrality
since one has to include the contribution of the particle-particle
interaction with the same prescription.\\
With the notations of subsection III.A, one has
\beq
E_{pp}(\epsilon)=E_{\bm{r}}+E_{\bm{\kappa}\neq\bm{0}}+\frac{q^{2}\sqrt{\pi}}{L^{2}}\sum_{i=1}^{N}\sum_{j=1,j\neq i}^{N}\int_{\epsilon}^{\alpha^{2}}\frac{dt}{t^{\frac{3}{2}}}e^{-z_{ij}^{2}t}+E_{pp}^{Self}(\epsilon)
\eeq
Computing the integral in the right handed side of Eq.(A.10), one finds 
\beq
\begin{array}{ll}
\displaystyle E_{pp}(\epsilon)&\displaystyle =E_{\bm{r}}+E_{\bm{\kappa}\neq\bm{0}}+E_{\bm{\kappa}=\bm{0}}\\
&\\
&\displaystyle+\frac{2 q^{2}\pi}{L^{2}}\sum_{i=1}^{N}\sum_{j=1,j\neq i}^{N}\mbox{\LARGE{(}}\frac{e^{-z_{ij}^{2}\epsilon}}{\sqrt{\epsilon\pi}}+\mid z_{ij}\mid\mbox{erf}(\sqrt{\epsilon}\mid z_{ij}\mid)\mbox{\LARGE{)}}+E_{pp}^{Self}(\epsilon)
\end{array}
\eeq
The particle-particle contribution is then separated into the converging
contribution given in Subsection III.A and a diverging contribution as 
\beq
E_{pp}(\epsilon)=E_{pp}^{(cv)}+E_{pp}^{(dv)}(\epsilon)
\eeq
with
\beq
E_{pp}^{(dv)}(\epsilon)=\frac{N^{2}q^{2}}{L^{2}}\frac{\sqrt{\pi}}{\sqrt{\epsilon}}+\frac{q^{2}\sqrt{\pi}}{L^{2}}\mbox{\LARGE{(}}\sum_{i=1}^{N}\sum_{j=1,j\neq i}^{N}z_{ij}^{2}\mbox{\LARGE{)}}\sqrt{\epsilon}+o(\epsilon^{\frac{3}{2}})
\eeq
Thus, adding the contributions given by Eqs.(A.5), (A.9), (A.11) and
(A.13), and letting $\epsilon\rightarrow 0$, Eq.(A.2) becomes 
\beq
E=E_{pp}^{(cv)}+\frac{\pi}{2}\frac{N^{2}q^{2}}{L^{2}}h=E_{pp}^{(cv)}+2\pi\sigma^{2}L^{2}h=E_{pp}^{(cv)}+E_{W}
\eeq
The contributions of the charged surfaces to the energy when the
particles are confined between surfaces is simply a term that is not
depending on the position of the particles. During the sampling of the
phase space of the system described in section II, especially when the
energy difference for a trial move of a particle is computed, the term
added in Eq.(A.14), $E_{W}$ is constant as long as $h$, $L$ and $N$ are
constants. This constant is included in the results given in TABLE III
and IV, and Figures 2(a-d). $E_{W}$ is considered as a contribution to
interlayer energies.\\   
On the other hand, $E_{pp}^{(cv)}$ is evaluated using the Ewald
quasi-2D, Hautman-Klein or Lekner-cyclic methods.

\newpage
\listoftables
\normalsize{\bf TABLE I:} Geometrical and simulation parameters of the
different Monte Carlo samplings. $L$ is the spatial periodicity, $h$ the
distance between the layers, $N$ the number of point ions in the
simulation, $\sigma$ the uniform surface charge density used to achieve
electroneutrality, $t_{eq}$ the number of Monte Carlo cycles done in
each Runs for equilibration, $t_{av}$ the number of Monte Carlo cycles
to accumulate thermodynamical average, $\beta u_{0}$ the Mandelung
energy computed for a square bidimensional lattice. The preferential
structure is the structure found in the Monte carlo study of the
solid-phase diagram of the classical bilayer crystal done in ref.[31].\\  

\normalsize{\bf TABLE II:} Estimation of the accuracy of the sampling of
the phase space by the three different methods. $\beta \Delta
E_{pp}^{(cv)}$ is the energy difference computed with the three methods
of the first trial move of the first MC-cycle from the same initial
conditions and with the same random numbers. ACC($t=1$) is the number of
accepted trial moves after the first MC-cycle and ACC($t=100$) the
number of accepted trial moves after the first hundred MC-cycles (see
also Figures 2).\\

\normalsize{\bf TABLE III:} Averages energies computed with the Monte
Carlo algorithm for Runs a and b ($h=1.0$). $\beta U/N$ is the average
total energy per particle, $\beta E_{inter}/N$ and $\beta E_{intra}/N$
are respectively the average interlayer and intralayer energy per
particle, $\beta\sigma_{U}/N$ an estimation of the statistical
fluctuation $\sigma_{U}$ of the total energy and $\delta t$ is a crude
estimation of the average CPU-time in seconds per MC-cycle
($\beta=1/kT$, $T$ temperature).\\   

\normalsize{\bf TABLE IV:} Same as TABLE III, but for Runs c, d, e, e'
and f ($h=4.0$).\\

\newpage
\listoffigures

\normalsize{\bf Figure 1:} Representation of the bilayer Wigner
cristal. Periodic boundary conditions with a spatial periodicity $L$ are
not represented. This representation is a snapshot of an instantaneous
configuration of a system of 512 point ions with the geometrical
parameters of the Runs e and e' given in TABLE I.\\

\normalsize{\bf Figure 2:} Instantaneous energy per particle for the
first 500 Monte-Carlo cycles computed with different methods but with
exactly the same initial configuration and with the same sequence of
random numbers. The parameters of these computations are: $h=1.0$,
$N=512$, $L=28.36$ (Run a) (see also TABLE II). (a) Comparison of the
sampling done with the Ewald quasi-2D and Hautman-Klein method. (b)
Comparison between the Ewald quasi-2D and Lekner-cyclic ($n_{c}=10$,
$n_{K}=3$) methods.\\
     
\normalsize{\bf Figure 2:} (c) Comparison between the Ewald quasi-2D and
Lekner-cyclic ($n_{c}=25$, $n_{K}=3$) methods. (d) Comparison between the
Ewald quasi-2D and Lekner-cyclic ($n_{c}=50$, $n_{K}=3$) methods.\\

\normalsize{\bf Figure 3:} Representation of the intralayer $g_{11}(s)$
and interlayer $g_{12}(s)$ pair distributions functions obtained with
the Ewald quasi-2D and Hautman Klein methods for Runs a and b
($h=1.0$). (a) $g_{11}(s)$ from Runs a; (b) $g_{12}(s)$ from Runs a; (c)
$g_{11}(s)$ from Runs b; (d) $g_{12}(s)$ from Runs b.\\

\normalsize{\bf Figure 4:} Same as the Figure 3, but for the Ewald
quasi-2D and Lekner-cyclic ($n_{c}=10$, $n_{K}=3$) methods for Runs a and
b ($h=1.0$). (a) $g_{11}(s)$ from Runs a; (b) $g_{12}(s)$ from Runs a;
(c) $g_{11}(s)$ from Runs b; (d) $g_{12}(s)$ from Runs b.\\ 

\normalsize{\bf Figure 5:} Same as the Figure 3, but for the Ewald
quasi-2D and Lekner-cyclic ($n_{c}=25$, $n_{K}=3$) methods for Runs a and
b ($h=1.0$). (a) $g_{11}(s)$ from Runs a; (b) $g_{12}(s)$ from Runs a;
(c) $g_{11}(s)$ from Runs b; (d) $g_{12}(s)$ from Runs b.\\ 

\normalsize{\bf Figure 6:} Same as the Figure 3, but for the Ewald
quasi-2D and Lekner-cyclic ($n_{c}=40$, $n_{K}=2$) methods for Runs a and
b ($h=1.0$). (a) $g_{11}(s)$ from Runs a; (b) $g_{12}(s)$ from Runs a;
(c) $g_{11}(s)$ from Runs b; (d) $g_{12}(s)$ from Runs b.\\ 

\normalsize{\bf Figure 7:} Same as Figure 3, but for the Ewald quasi-2D
and Hautman Klein methods for Runs c to f ($h=4.0$). (a) $g_{11}(s)$
from Runs c; (b) $g_{12}(s)$ from Runs c ($N=128$); (c) $g_{11}(s)$ from
Runs d ; (d) $g_{12}(s)$ from Runs d ($N=338$); (e) $g_{11}(s)$ from
Runs e and e'; (f) $g_{12}(s)$ from Runs e and e' ($N=512$); (g)
$g_{11}(s)$ from Runs f ; (h) $g_{12}(s)$ from Runs f ($N=968$).\\  

\normalsize{\bf Figure 8:} Same as Figure 3, but for the Ewald quasi-2D
and Lekner-cyclic ($n_{c}=10$, $n_{K}=3$) methods for Runs c to f'
($h=4.0$). (a) $g_{11}(s)$ from Runs c; (b) $g_{12}(s)$ from Runs c
($N=128$); (c) $g_{11}(s)$ from Runs d ; (d) $g_{12}(s)$ from Runs d
($N=338$); (e) $g_{11}(s)$ from Runs e and e' ; (f) $g_{12}(s)$ from
Runs e and e'  ($N=512$); (g) $g_{11}(s)$ from Runs f ; (h) $g_{12}(s)$
from Runs f ($N=968$).\\   

\normalsize{\bf Figure 9:} Same as Figure 3, but for the Ewald quasi-2D
and Lekner-cyclic ($n_{c}=25$, $n_{K}=3$) methods for Runs c to f'
($h=4.0$). (a) $g_{11}(s)$ from Runs c; (b) $g_{12}(s)$ from Runs c
($N=128$); (c) $g_{11}(s)$ from Runs d ; (d) $g_{12}(s)$ from Runs d
($N=338$); (e) $g_{11}(s)$ from Runs e and e' ; (f) $g_{12}(s)$ from
Runs e and e' ($N=512$); (g) $g_{11}(s)$ from Runs f ; (h) $g_{12}(s)$
from Runs f ($N=968$).\\  

\newpage

\begin{figure}
\begin{center}
\epsfysize=12.truein
\vbox{\vskip -0.6truein \hskip -0.3truein
\epsffile{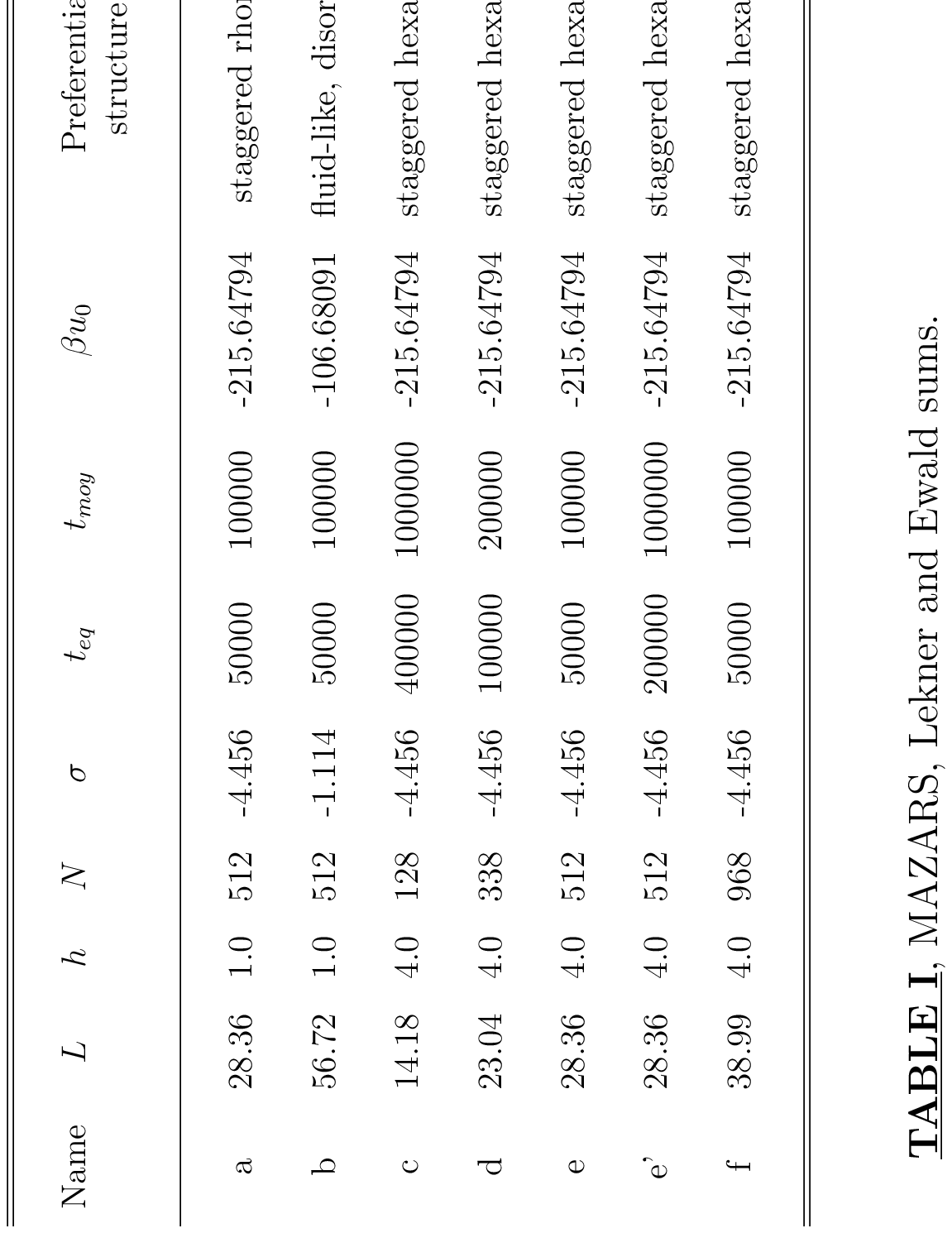}}
\end{center}
\end{figure}

\newpage

\begin{center}
\begin{tabular}{ccccc}
\hline
\hline
&&&&\\
Method & Parameters & $\beta \Delta E_{pp}^{(cv)} $ & ACC & ACC  \\
&of the method & & $(t=1)$  & $(t=100)$ \\
&&&&\\
\hline
&&&&\\
Ewald quasi-2D & $\alpha=0.3$, $\kappa\times\kappa= 8\times 8$ & -0.55046(0) & 243 & 23139\\        
&&&&\\
\hline
&&&&\\
Hautman-Klein & $\alpha=0.3$, $\kappa\times\kappa= 8\times 8$ &  &&\\
	      &		$M=1$		        &  -0.55062(7) & 243 & 23171\\
	      &		$M=2$			&  -0.55063(1) & 243 & 23171\\
	      &		$M=3$			&  -0.55063(2) & 243 & 23171\\
&&&&\\
\hline
&&&&\\
Lekner-cyclic & $n_{c}=5$, $n_{K}=1$		&  -1.37408(7) & 348 & 24623\\
	      & $n_{c}=5$, $n_{K}=2$		&  -1.37404(2) & 348 & 24623\\
	      & $n_{c}=5$, $n_{K}=3$		&  -1.37404(1) & 348 & 24623\\
	      & $n_{c}=5$, $n_{K}=4$		&  -1.37404(1) & 348 & 24623\\
	      & $n_{c}=10$, $n_{K}=1$		&  -0.77481(0) & 263 & 24775\\
	      & $n_{c}=10$, $n_{K}=2$		&  -0.77476(4) & 263 & 24775\\
	      & $n_{c}=10$, $n_{K}=3$		&  -0.77476(4) & 263 & 24775\\
	      & $n_{c}=10$, $n_{K}=4$		&  -0.77476(4) & 263 & 24775\\
	      & $n_{c}=15$, $n_{K}=1$		&  -0.51255(9) & 244 & 23601\\
	      & $n_{c}=15$, $n_{K}=2$		&  -0.51251(3) & 244 & 23597\\
	      & $n_{c}=15$, $n_{K}=3$		&  -0.51251(3) & 244 & 23597\\
	      & $n_{c}=15$, $n_{K}=4$		&  -0.51251(3) & 244 & 23597\\
	      & $n_{c}=25$, $n_{K}=1$		&  -0.54498(5) & 243 & 23132\\
	      & $n_{c}=25$, $n_{K}=2$		&  -0.54493(9) & 243 & 23132\\
	      & $n_{c}=25$, $n_{K}=3$		&  -0.54493(9) & 243 & 23132\\
	      & $n_{c}=40$, $n_{K}=1$		&  -0.55093(7) & 243 & 23172\\
	      & $n_{c}=40$, $n_{K}=2$		&  -0.55089(1) & 243 & 23147\\
	      & $n_{c}=40$, $n_{K}=3$		&  -0.55089(1) & 243 & 23147\\
	      & $n_{c}=45$, $n_{K}=1$		&  -0.55075(8) & 243 & 23172\\
	      & $n_{c}=45$, $n_{K}=2$		&  -0.55071(1) & 243 & 23172\\
	      & $n_{c}=45$, $n_{K}=3$		&  -0.55071(2) & 243 & 23172\\
&&&&\\ 
\hline
\hline
\end{tabular}
\end{center}
\begin{center}
{\begin{quote}\item[\large\underline{\bf{TABLE II}}, MAZARS, Lekner and
Ewald sums.]\end{quote}} 
\end{center}

\newpage

\begin{figure}
\begin{center}
\epsfysize=12.truein
\vbox{\vskip -0.6truein \hskip -0.3truein 
\epsffile{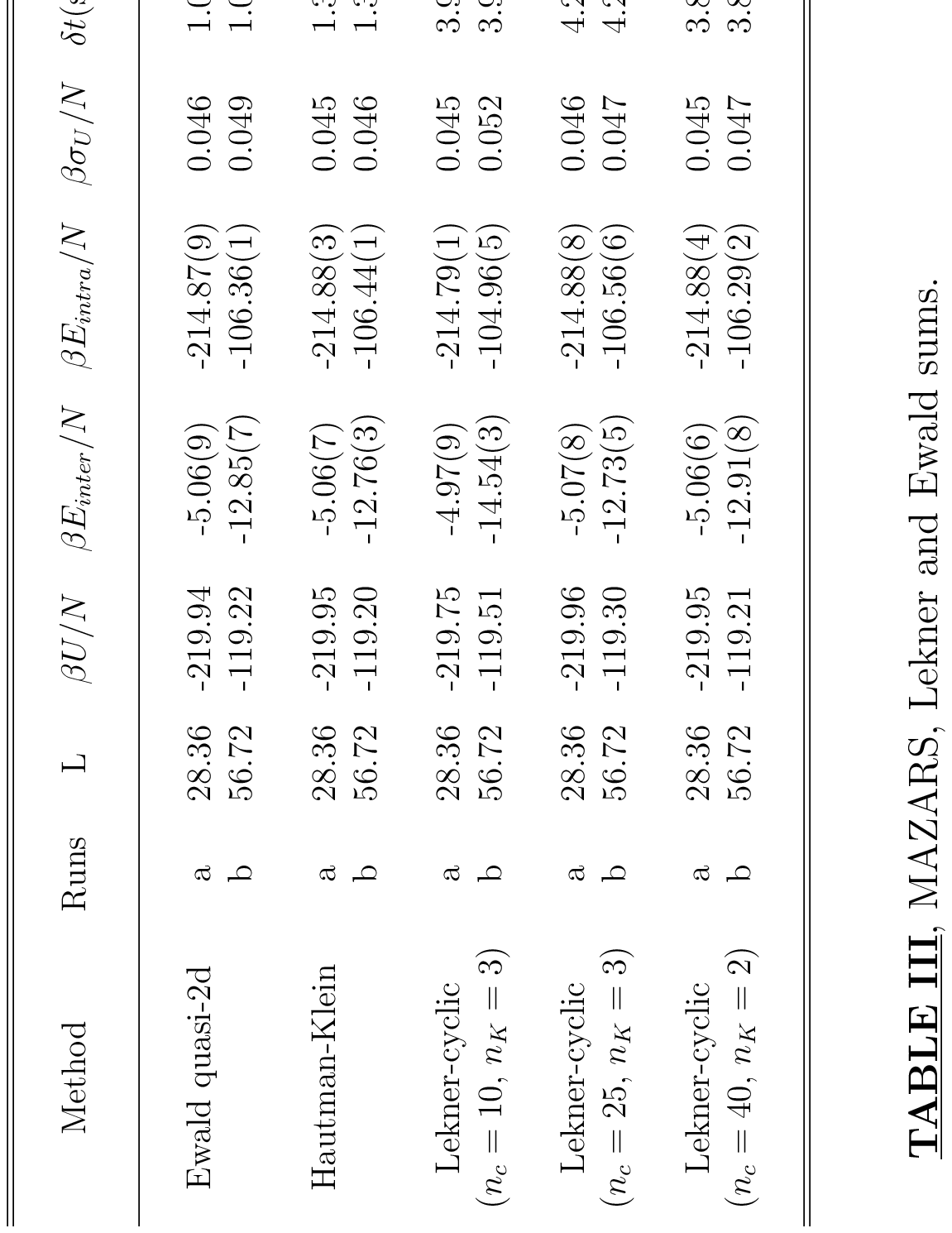}}
\end{center}
\end{figure}

\newpage

\begin{figure}
\begin{center}
\epsfysize=12.truein
\vbox{\vskip -0.6truein \hskip -0.3truein
\epsffile{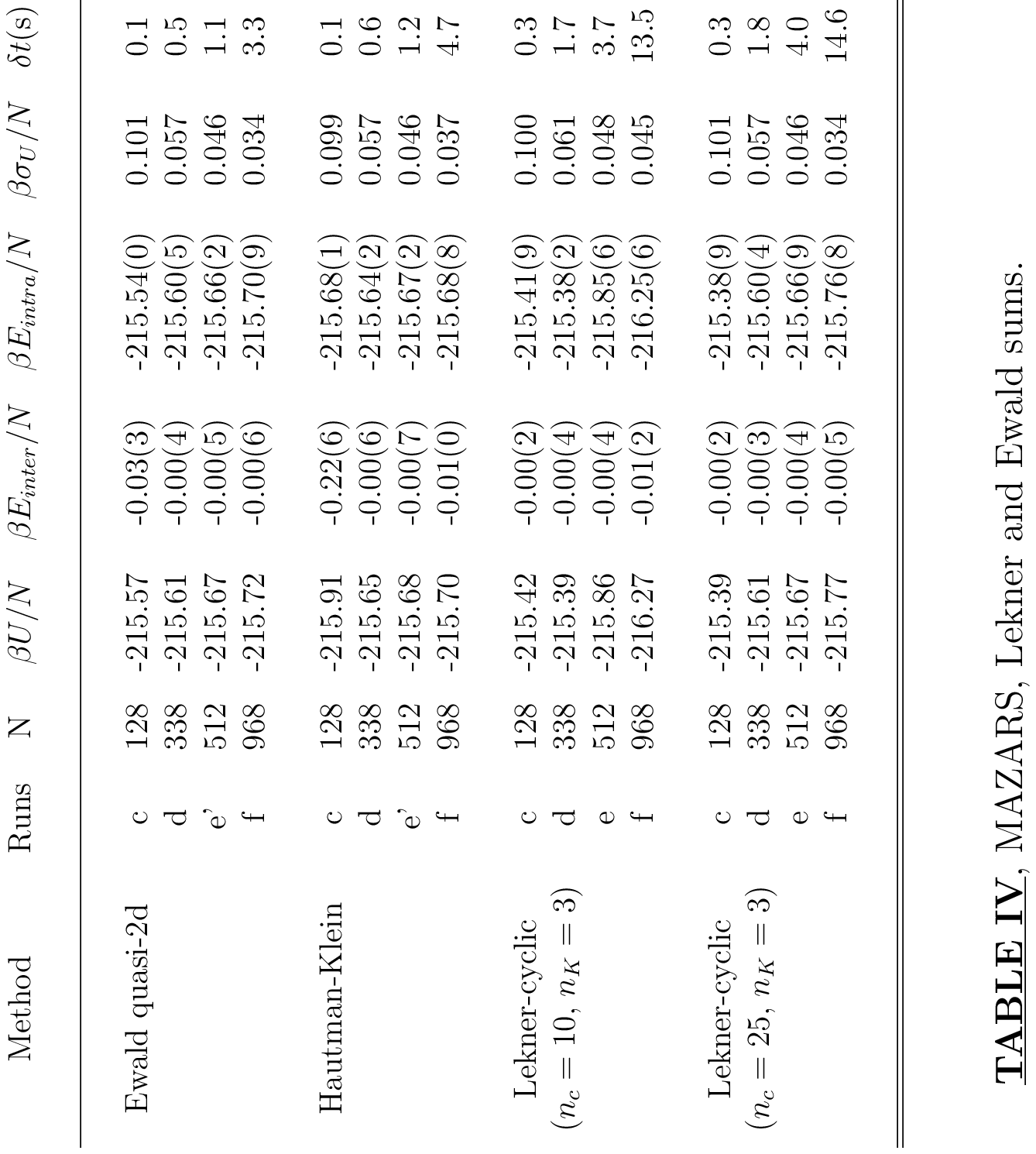}}
\end{center}
\end{figure}

\newpage

\begin{figure}
\begin{center}
\epsfysize=1.8truein
\vbox{\vskip 0.truein \hskip -0.3truein
\epsffile{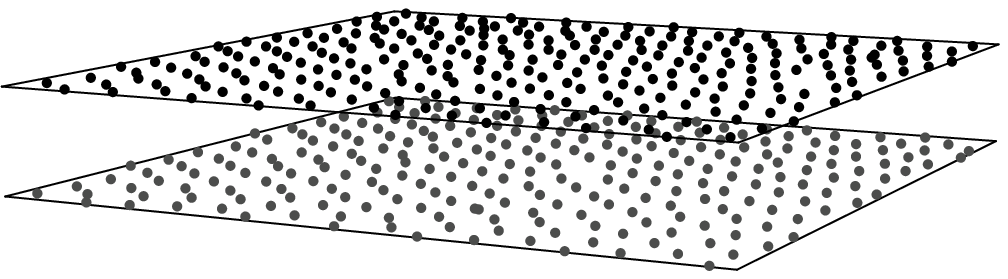}}
\end{center}
{\begin{quote}\item[\large\underline{\bf{Figure 1}}, MAZARS, Lekner and
Ewald sums.]\end{quote}} 
\end{figure}

\newpage

\begin{figure}
\begin{center}
\epsfysize=6.truein
\vbox{\vskip 0.truein \hskip -0.3truein
\epsffile{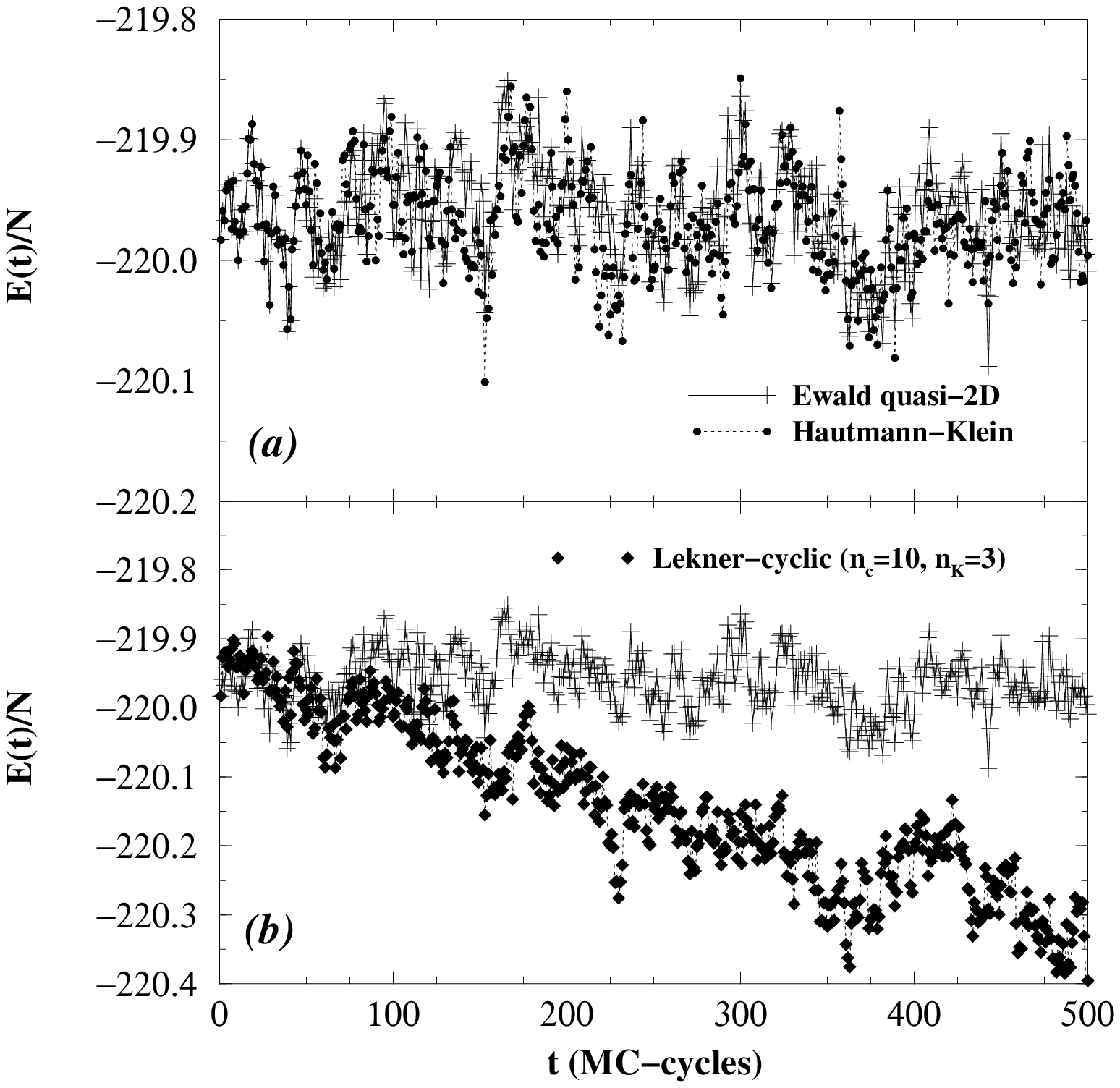}}
\end{center}
{\begin{quote}\item[\large\underline{\bf{Figure 2(a),(b)}}, MAZARS,
Lekner and Ewald sums.]\end{quote}} 
\end{figure}

\newpage

\begin{figure}
\begin{center}
\epsfysize=6.truein
\vbox{\vskip 0.truein \hskip -0.3truein
\epsffile{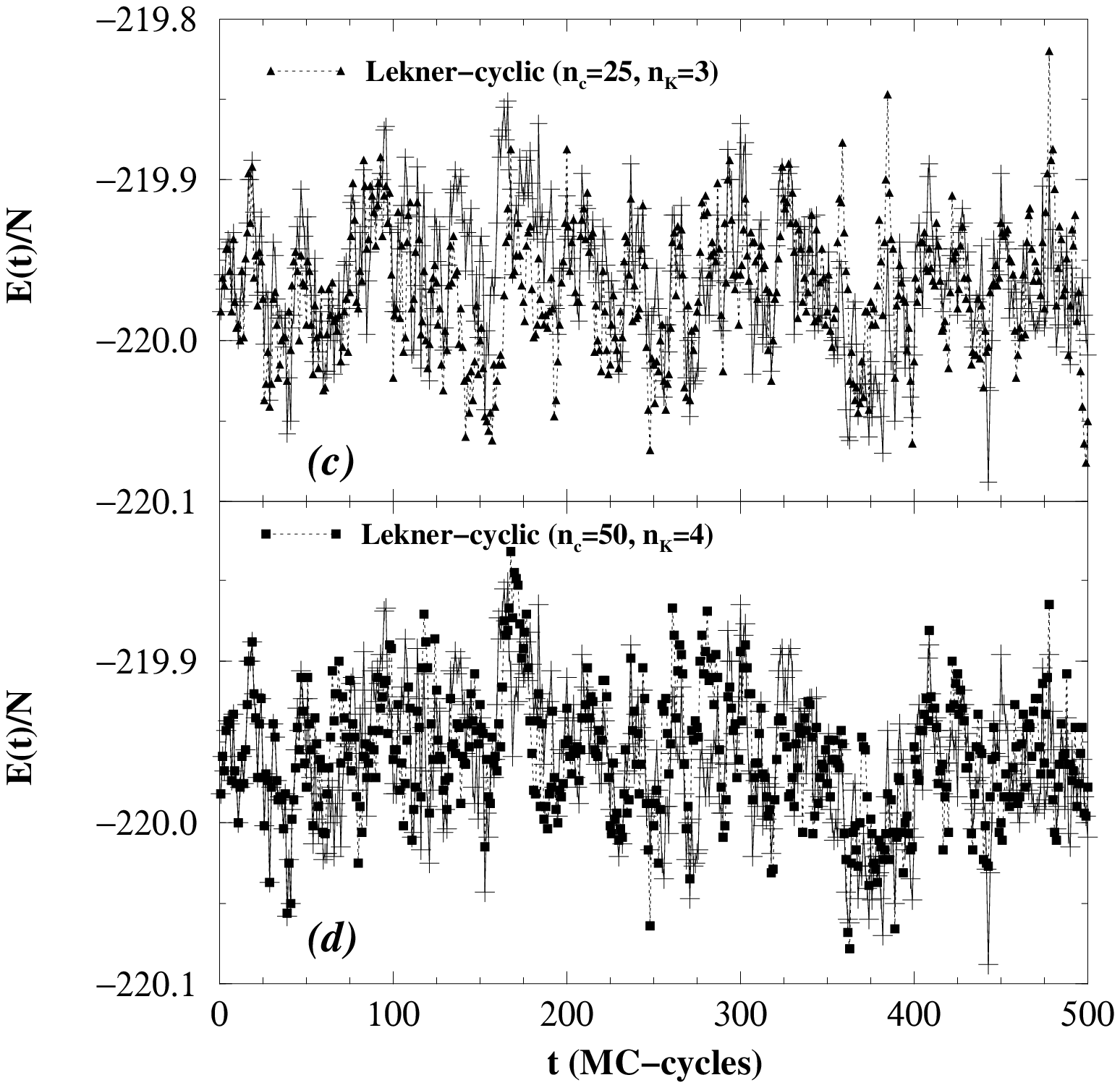}}
\end{center}
{\begin{quote}\item[\large\underline{\bf{Figure 2(c),(d)}}, MAZARS,
Lekner and Ewald sums.]\end{quote}} 
\end{figure}

\newpage
\begin{figure}
\begin{center}
\epsfysize=5.truein
\vbox{\vskip 0.truein \hskip -0.3truein
\epsffile{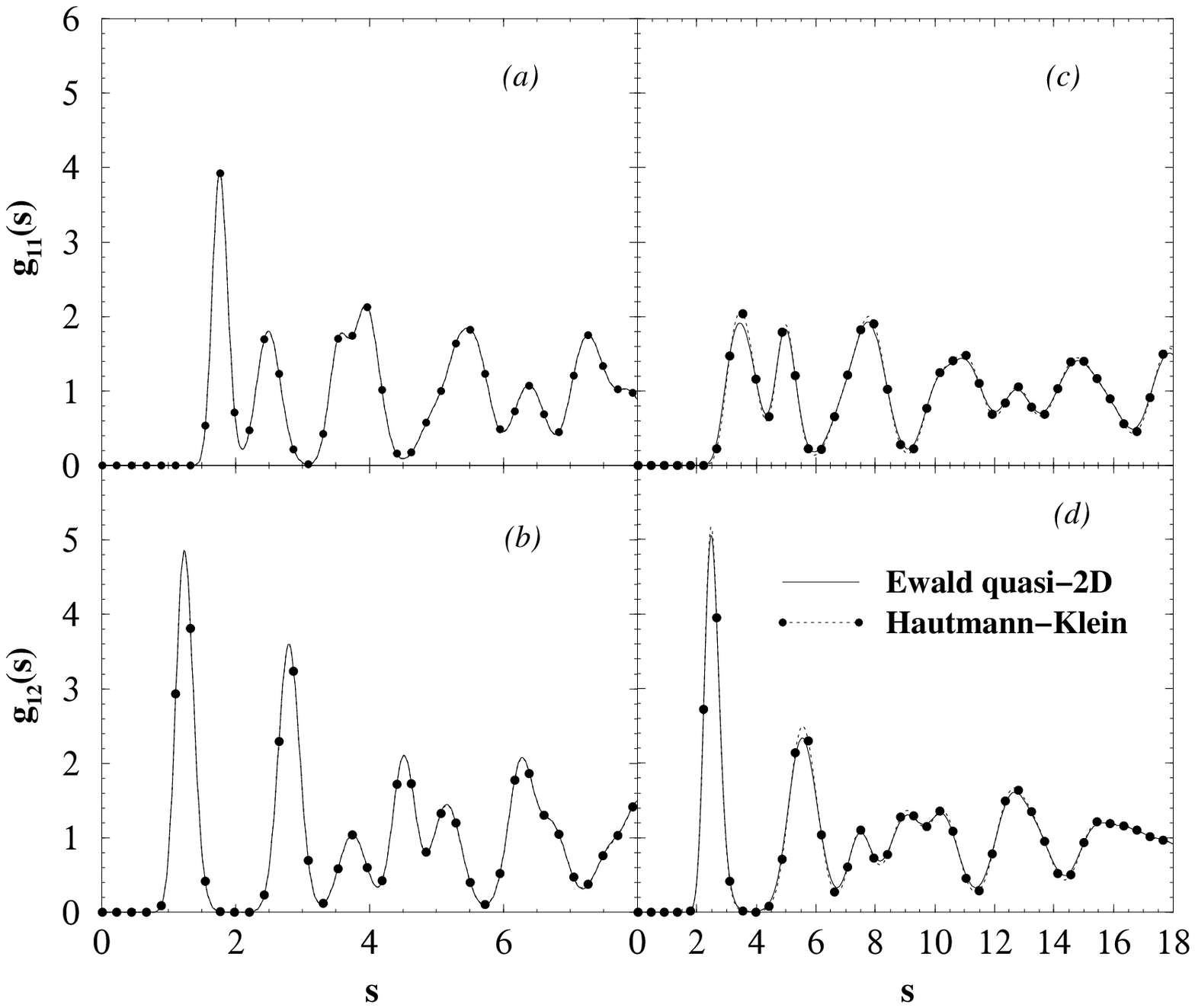}}
\end{center}
{\begin{quote}\item[\large\underline{\bf{Figure 3 (a-d)}},
MAZARS, Lekner and Ewald sums.]\end{quote}} 
\end{figure}

\newpage
\begin{figure}
\begin{center}
\epsfysize=5.truein
\vbox{\vskip 0.truein \hskip -0.3truein
\epsffile{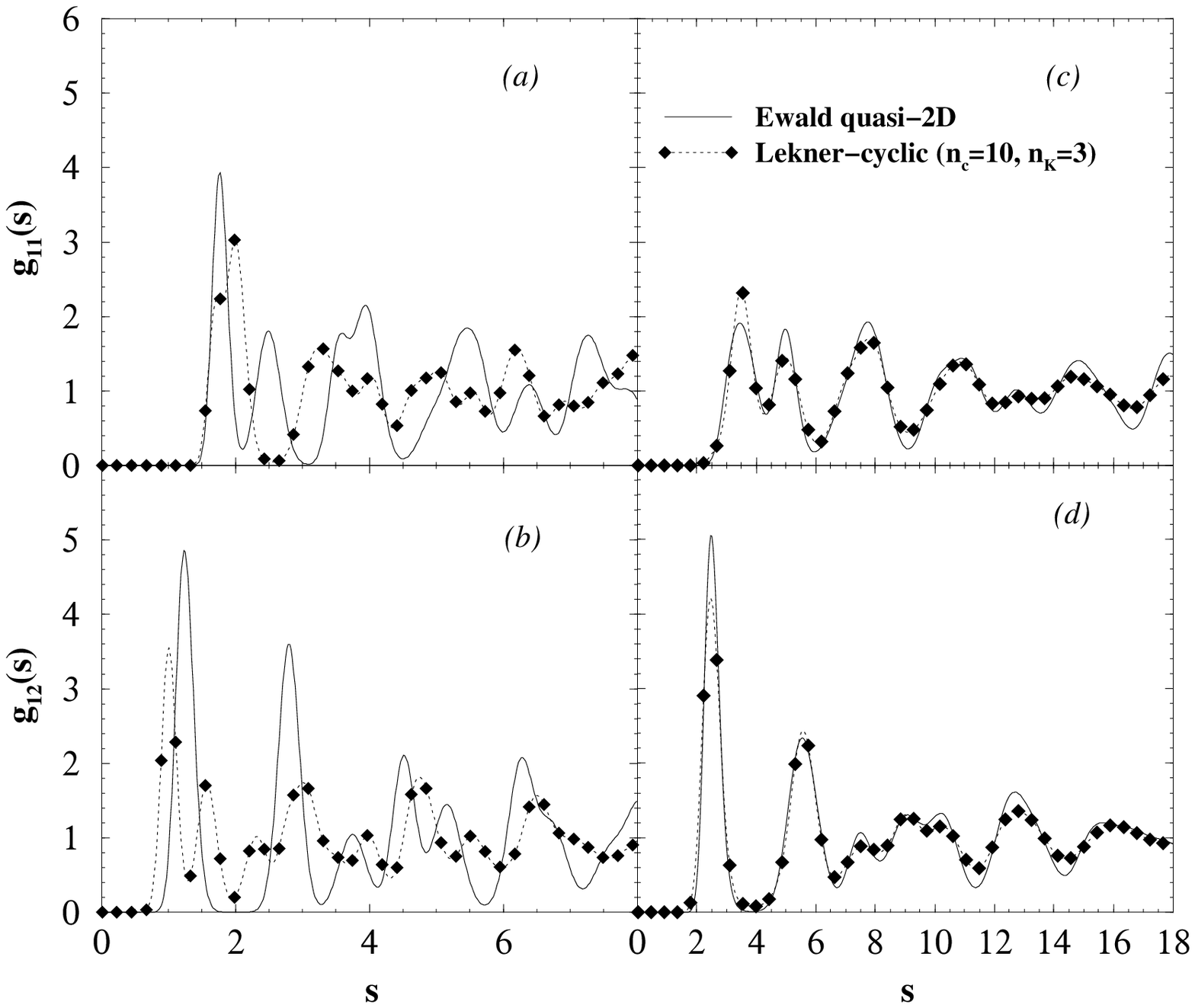}}
\end{center}
{\begin{quote}\item[\large\underline{\bf{Figure 4 (a-d)}},
MAZARS, Lekner and Ewald sums.]\end{quote}} 
\end{figure}

\newpage
\begin{figure}
\begin{center}
\epsfysize=5.truein
\vbox{\vskip 0.truein \hskip -0.3truein
\epsffile{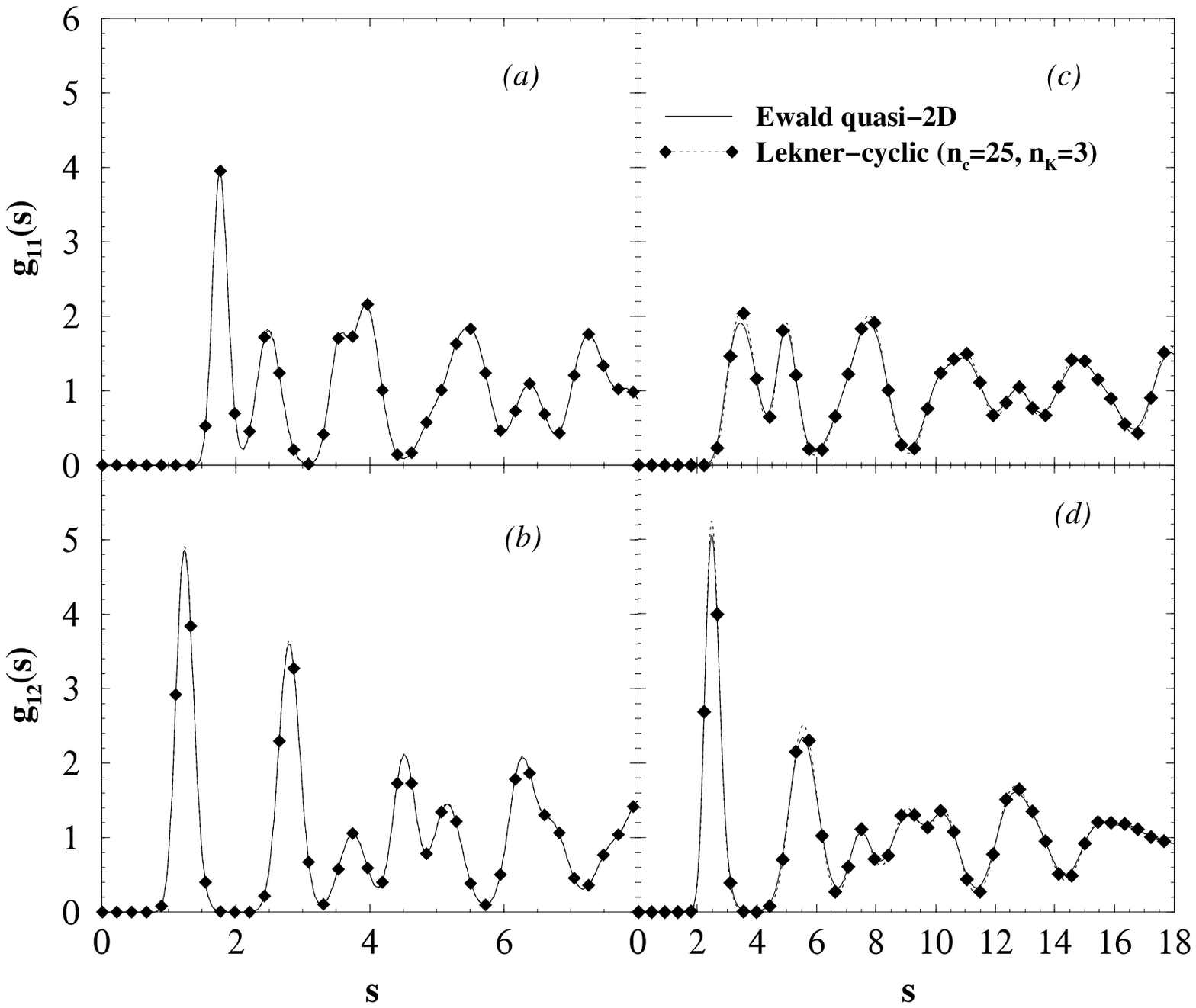}}
\end{center}
{\begin{quote}\item[\large\underline{\bf{Figure 5 (a-d)}},
MAZARS, Lekner and Ewald sums.]\end{quote}} 
\end{figure}

\newpage
\begin{figure}
\begin{center}
\epsfysize=5.truein
\vbox{\vskip 0.truein \hskip -0.3truein
\epsffile{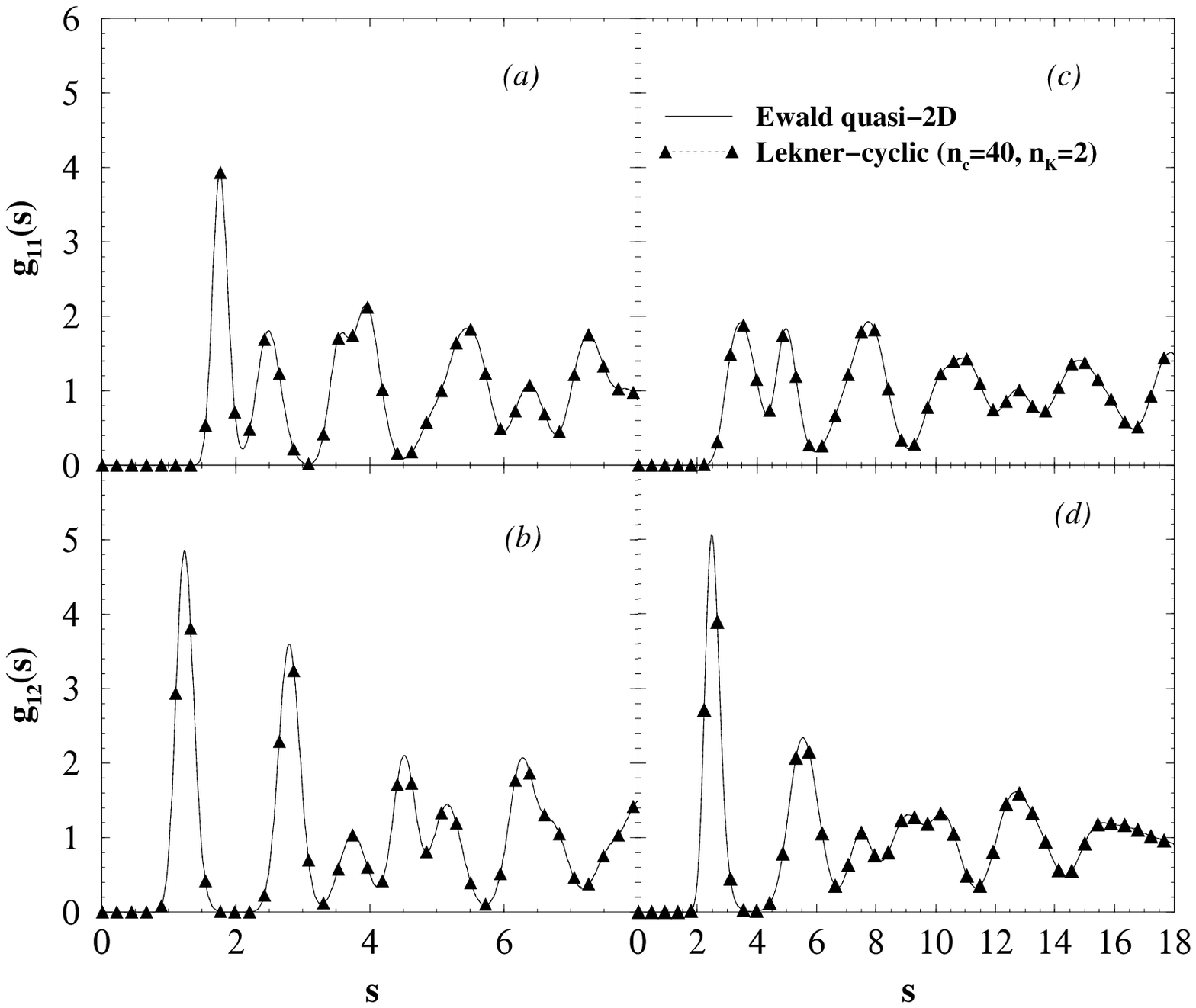}}
\end{center}
{\begin{quote}\item[\large\underline{\bf{Figure 6 (a-d)}},
MAZARS, Lekner and Ewald sums.]\end{quote}} 
\end{figure}

\newpage
\begin{figure}
\epsfysize=9.truein
\vbox{\vskip -0.3truein \hskip -0.truein \epsffile{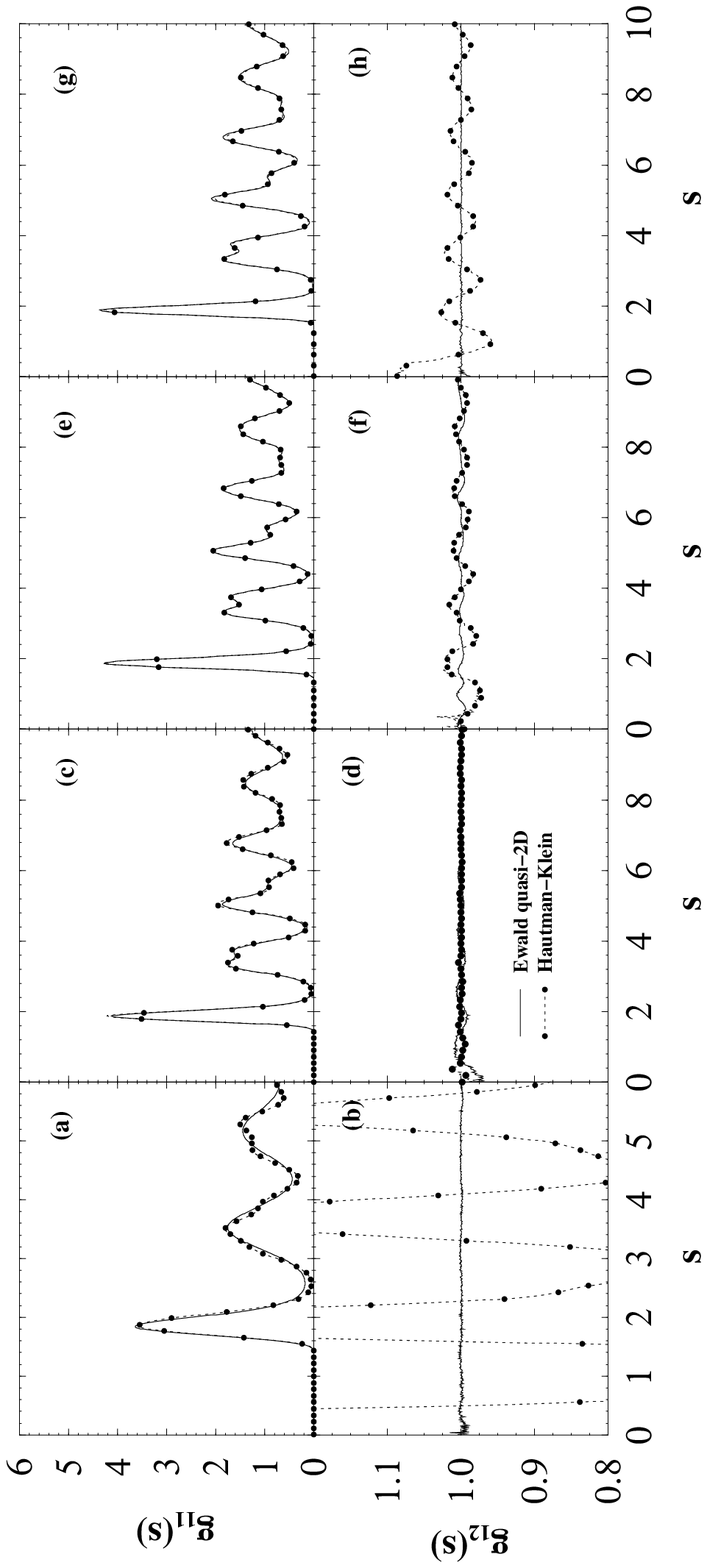}}
{\begin{quote}\item[\large\underline{\bf{Figure 7
(a-h)}}, MAZARS, Lekner and Ewald
sums. ]\end{quote}} 
\end{figure}

\newpage
\begin{figure}
\epsfysize=9.truein
\vbox{\vskip -0.3truein \hskip -0.truein \epsffile{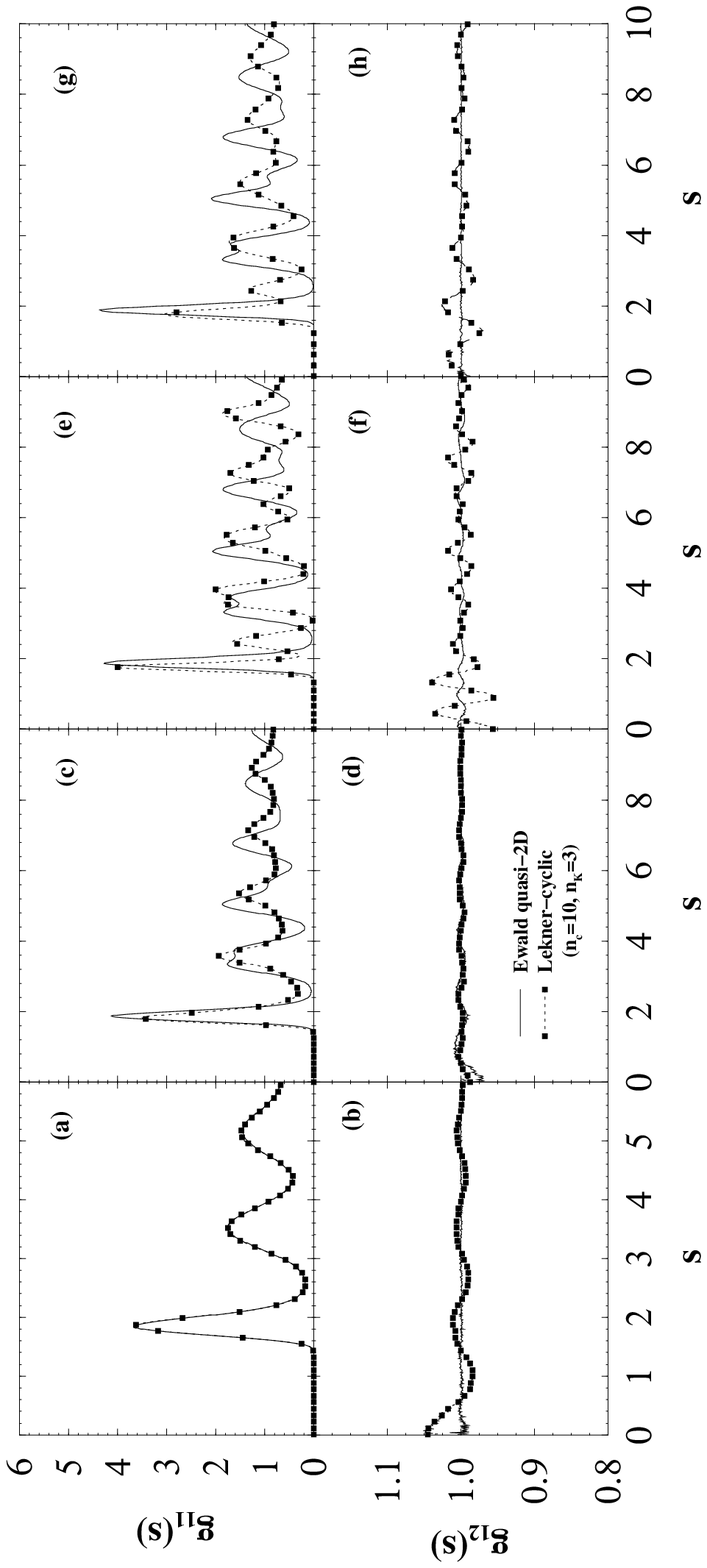}}
{\begin{quote}\item[\large\underline{\bf{Figure 8
(a-h)}}, MAZARS, Lekner and Ewald 
sums.]\end{quote}} 
\end{figure}

\newpage
\begin{figure}
\epsfysize=9.truein
\vbox{\vskip -0.3truein \hskip -0.truein \epsffile{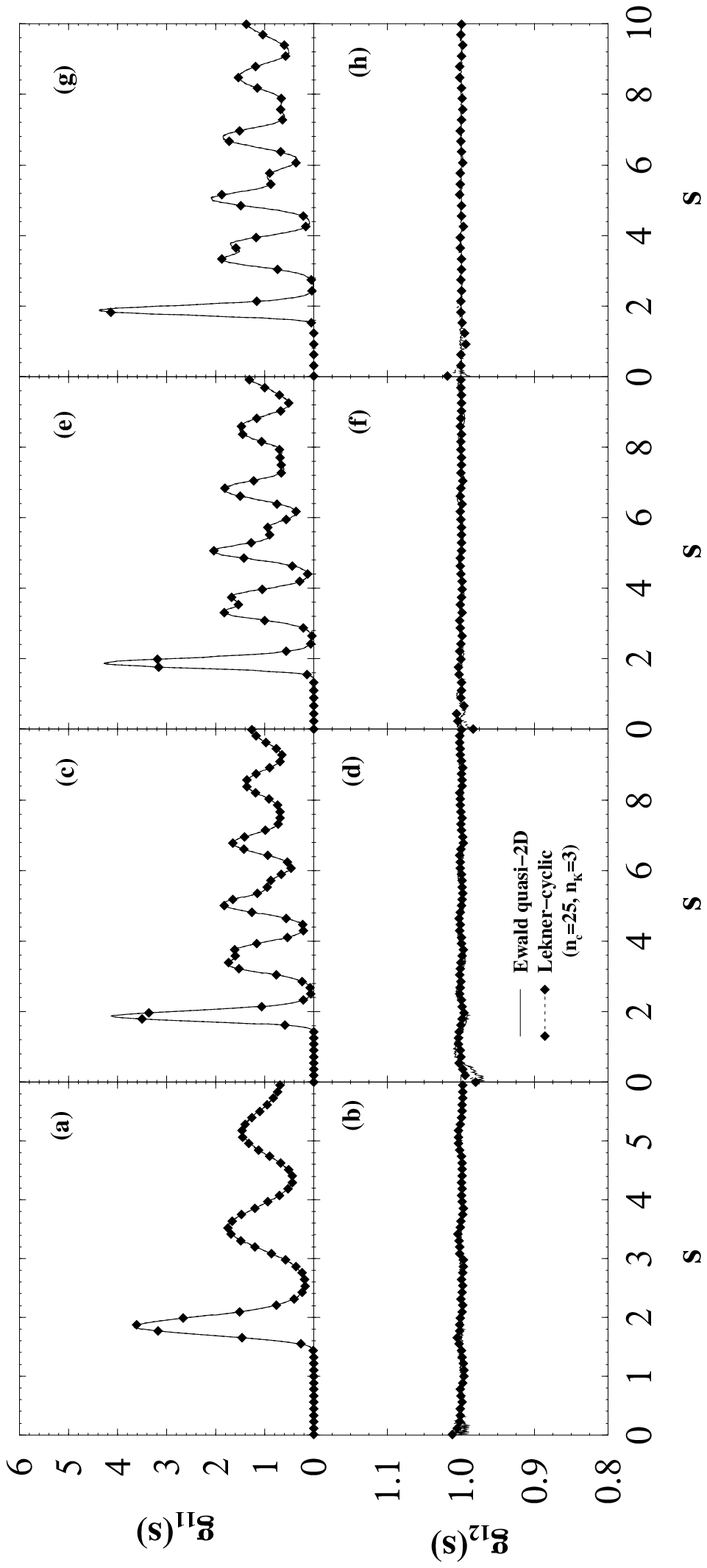}}
{\begin{quote}\item[\large\underline{\bf{Figure 9
(a-h)}}, MAZARS, Lekner and Ewald
sums.]\end{quote}} 
\end{figure}

\end{document}